\documentstyle[12pt,epsfig]{article}
\newlength{\dinwidth}
\newlength{\dinmargin}
\newlength{\gap}
\newcommand{\kz}{\mbox{$K^{0}$}}
\newcommand{\kzs}{\mbox{$K^{0}_s$}}

\newcommand{\hpm}{\mbox{$h^{\pm}$}}
\newcommand{\pT}{\mbox{$p_T$}}
\newcommand{\pTh}{\mbox{$p_T$($h^\pm$)}}
\newcommand{\pTk}{\mbox{$p_T$($K^0$)}}

\newcommand{\eT}{\mbox{$E_T$}}
\newcommand{\eTj}{\mbox{$E_T$(jet)}}
\newcommand{\eTrec}{\mbox{$E_T^{\;rec}$}}
\newcommand{\etaj}{\mbox{$\eta$(jet)}}
\newcommand{\etah}{\mbox{$\eta$($h^\pm$)}}
\newcommand{\etak}{\mbox{$\eta$($K^0$)}}
\newcommand{\etaks}{\mbox{$\eta$($K^0_s$)}}
\newcommand{\pB}{\mbox{\bf p}} 
\newcommand{\nB}{\mbox{\bf n}} 
 
\newcommand{\xgO}{\mbox{$x_\gamma^{\mbox{\tiny OBS}}$}}

\newcounter{ccc}
\newenvironment{numlis}{ 
\begin{list}{(\arabic{ccc})}{\usecounter{ccc}\setlength{\itemsep}{1ex}
\setlength{\parsep}{0ex}
\setlength{\topsep}{0ex}}}{\end{list}}
\newcounter{ccd}
\newenvironment{romlis}{ 
\begin{list}{(\roman{ccd})}{\usecounter{ccd}\setlength{\itemsep}{1ex}
\setlength{\parsep}{0ex}
\setlength{\topsep}{0ex}}}{\end{list}}
\begin{document}
\thispagestyle{empty}
\settowidth{\gap}{\large Submitted to } 


\title{\bf Charged particles and neutral kaons 
in photoproduced jets at HERA\\[2cm]}
\author{ZEUS Collaboration\\[40mm]}
\date{}
\maketitle
\vspace{2cm}
\thispagestyle{empty}

\begin{abstract} 
Charged particles ($h^\pm$) and \kz\ mesons have been studied in 
photoproduced events containing at least one jet of $E_T > 8$ GeV
in a pseudorapidity interval (--0.5, 0.5) in the ZEUS laboratory frame. 
Distributions are presented in terms of transverse momentum, pseudorapidity 
and distance of the particle from the axis of a jet.  The properties 
of \hpm\ within the jet are described  well using the standard 
settings of PYTHIA, but the use of the multiparton interaction option 
improves the description outside the jets.  A reasonable overall 
description of the \kz\ behaviour is possible  with PYTHIA using 
a reduced value of the strangeness suppression parameter.  
The numbers of $h^\pm$ and \kz\ within a jet as defined above are measured to be 
$3.25\pm0.02\pm0.28$ and $0.431\pm0.013\pm0.088$ respectively.    
Fragmentation functions are presented for $h^\pm$ and \kz\ in 
photoproduced jets; agreement is found with calculations of 
Binnewies et al.\  and, at higher momenta, with 
$p\bar p$ scattering and with standard PYTHIA.
Fragmentation functions in direct photoproduced events are extracted, and 
at higher momenta give good agreement with data from related processes 
in $e^+e^-$ annihilation and deep inelastic $ep$ scattering. 
\end{abstract}
\vspace{-23.4cm}{\noindent DESY-97-229\newline November 1997}
\thispagestyle{empty}

\newpage 
\parindent0.cm                                                                                     
\parskip0.3cm plus0.05cm minus0.05cm                                                               
\def\3{\ss}                                                                                        
\pagenumbering{Roman}                                                                              
\newcommand{\address}{}
\newcommand{\author}{}                                                   %
\begin{center}                                                                                     
{                      \Large  The ZEUS Collaboration              }                               
\end{center}                                                                                       
  J.~Breitweg,                                                                                     
  M.~Derrick,                                                                                      
  D.~Krakauer,                                                                                     
  S.~Magill,                                                                                       
  D.~Mikunas,                                                                                      
  B.~Musgrave,                                                                                     
  J.~Repond,                                                                                       
  R.~Stanek,                                                                                       
  R.L.~Talaga,                                                                                     
  R.~Yoshida,                                                                                      
  H.~Zhang  \\                                                                                     
 {\it Argonne National Laboratory, Argonne, IL, USA}~$^{p}$
% ...
\newpage                                                                                           
% ...
$^{  23}$ now at ATLAS Collaboration, Univ. of Munich \\                                           
% ...
$^{  33}$ partially supported by the Foundation for German-Russian Collaboration                   
DFG-RFBR \\ \hspace*{3.5mm} (grant no. 436 RUS 113/248/3 and no. 436 RUS 113/248/2)\\              
% ...
                                                           %
                                                           %
\newpage   
                                                           %
                                                           %
\begin{tabular}[h]{rp{14cm}}                                                                       
$^{a}$ &  supported by the Natural Sciences and Engineering Research                               
          Council of Canada (NSERC)  \\                                                            
$^{b}$ &  supported by the FCAR of Qu\'ebec, Canada  \\                                            
$^{c}$ &  supported by the German Federal Ministry for Education and                               
          Science, Research and Technology (BMBF), under contract                                  
          numbers 057BN19P, 057FR19P, 057HH19P, 057HH29P, 057SI75I \\                              
$^{d}$ &  supported by the MINERVA Gesellschaft f\"ur Forschung GmbH,                              
          the German Israeli Foundation, and the U.S.-Israel Binational                            
          Science Foundation \\                                                                    
$^{e}$ &  supported by the German Israeli Foundation, and                                          
          by the Israel Science Foundation                                                         
  \\                                                                                               
$^{f}$ &  supported by the Italian National Institute for Nuclear Physics                          
          (INFN) \\                                                                                
$^{g}$ &  supported by the Japanese Ministry of Education, Science and                             
          Culture (the Monbusho) and its grants for Scientific Research \\                         
$^{h}$ &  supported by the Korean Ministry of Education and Korea Science                          
          and Engineering Foundation  \\                                                           
$^{i}$ &  supported by the Netherlands Foundation for Research on                                  
          Matter (FOM) \\                                                                          
$^{j}$ &  supported by the Polish State Committee for Scientific                                   
          Research, grant No.~115/E-343/SPUB/P03/002/97, 2P03B10512,                               
          2P03B10612, 2P03B14212, 2P03B10412 \\                                                    
$^{k}$ &  supported by the Polish State Committee for Scientific                                   
          Research (grant No. 2P03B08308) and Foundation for                                       
          Polish-German Collaboration  \\                                                          
$^{l}$ &  partially supported by the German Federal Ministry for                                   
          Education and Science, Research and Technology (BMBF)  \\                                
$^{m}$ &  supported by the Fund for Fundamental Research of Russian Ministry                       
          for Science and Edu\-cation and by the German Federal Ministry for                       
          Education and Science, Research and Technology (BMBF) \\                                 
$^{n}$ &  supported by the Spanish Ministry of Education                                           
          and Science through funds provided by CICYT \\                                           
$^{o}$ &  supported by the Particle Physics and                                                    
          Astronomy Research Council \\                                                            
$^{p}$ &  supported by the US Department of Energy \\                                              
$^{q}$ &  supported by the US National Science Foundation \\                                       
\end{tabular}                                                                                      
                                                           %
                                                           %

\newpage
\pagestyle{plain}
\pagenumbering{arabic} 
\setcounter{page}{1}
\parskip 3mm plus 2mm minus 2mm

\section{Introduction}
When a high energy photon interacts with a proton, a QCD-governed hard
scattering process may take place in which either the photon, or a parton 
within the photon, interacts with a parton in the proton. A signature for
such processes is high-\eT\ jets in the final state.  Both H1 and ZEUS at HERA 
have reported measurements of various aspects of jet photoproduction~[1--5].  
In general terms, the properties of particles in jets depend on the type of
leading parton  generated by the hard process, 
and on the process known as fragmentation (or hadronisation) 
by which the colour in the jet is neutralised. 
The main features of the fragmentation of a given leading parton
at a given energy are believed to be universal and independent of the type of 
initiating process.   
It is fragmentation, moreover,  rather than leading strange or 
charm quarks, which generates most of the strange particles that occur 
in jets~\cite{Cash}.                                        
At the phenomenological level, a longitudinal fragmentation function
$D(z)$ may be used to describe the observed momentum distribution of
a chosen type of particle along the jet axis, 
where $z$ is the fraction of the jet momentum taken by the particle. 
Experimentally, $D(z)$ functions 
have so far been obtained mainly from fits to $e^+ e^-$ collider data.  

In the present analysis, using data taken in $ep$ 
collisions with the ZEUS detector in 1994, we examine the inclusive properties 
of charged particles and \kz\  mesons produced 
in association with high-\eT\ photoproduced jets.   This work 
complements the studies of such particles  which have been performed in 
deep inelastic scattering (DIS) by the ZEUS and H1 
collaborations~[7-10].                                                                      
The results are compared with predictions obtained 
from the PYTHIA Monte Carlo generator, which incorporates
the Lund string model of particle fragmentation as implemented in 
JETSET~\cite{Pythia}.  As one possible improvement to the standard description, 
the suggestion is investigated that 
the particle distributions may be affected by an underlying event
structure due to multiparton scattering~\cite{ZEUSMI}. 

Several studies are also made in this analysis to test the universality of fragmentation. 
In deep inelastic scattering~\cite{H1DIS,ZEUSDIS,BrFr}, it has been found 
that the behaviour of charged particles in the current region of the
Breit frame is similar to that found in 
high energy quark jets in $e^+ e^-$ annihilation; however, there are 
indications~\cite{ZDISK,HDISK,E665}  that DIS  
may give jets with a lower strangeness content than expected from the standard
settings of PYTHIA. 
DELPHI~\cite{Delphi} has also reported results with a similar conclusion.
We investigate whether this applies also 
to jets produced using incoming photons that are quasi-real,
noting that inclusive cross sections of \kz\ in
photoproduction found by H1 are broadly consistent with PYTHIA 
predictions~\cite{H1K}. 
Comparison is also made with the fragmentation function calculations of 
Binnewies et al.~\cite{bin}, which are based on fits to the particle
content of jets at PEP and LEP~\cite{PEP}.  
To give a more immediate test of the universality of fragmentation in  
different types of process, we select a sample of events dominated by 
the direct photoproduction process (i.e.\ in which the photon interacts in a 
pointlike manner).  These data allow a 
comparison with $e^+e^-$ and DIS results without any intermediate
model or parameterisation. 

The structure of the paper is as follows.  After an account of the 
apparatus, the event selection and the selection of charged tracks, the 
procedure for reconstructing $K^0$ mesons is discussed. We describe the 
 Monte Carlo models used in simulating the data, and the 
more important sources of systematic error.  
Results are then given on the properties 
of charged particles and \kz\ mesons in photoproduced jets.  
Finally, we present a study of fragmentation functions for both 
charged particles and $K^0$ mesons, extracting
results for direct photoproduced events in order to make a general 
comparison with other types of experiment.

\section{Apparatus and running conditions} 

The data used in the present analysis were collected by the ZEUS detector 
at HERA. During 1994, HERA collided positrons with energy $E_e = 27.5$ GeV 
with protons of energy $E_p =820$ GeV, in 153 circulating 
bunches.  Additional unpaired positron (15) and proton (17) bunches enabled monitoring of 
beam related backgrounds.  The data sample used in this analysis corresponds to an integrated
luminosity of 2.6 pb$^{-1}$. The luminosity was measured by means of the positron-proton 
bremsstrahlung process $ep\to e\gamma p$, using a lead-scintillator calorimeter 
at $Z=-107$ m\footnote{The ZEUS
coordinates form a right-handed system with positive-$Z$ in the 
proton beam direction and a horizontal $X$-axis. The nominal 
interaction point is at $X = Y = Z = 0.$
Pseudorapidity $\eta$ is defined as $-\ln\tan\theta/2$, where
$\theta$ is the polar angle relative to the $Z$ direction. In the present 
analysis, $\eta$ is always defined in the laboratory frame, and the actual
interaction point of the event is taken into account.} 
which intercepts photons radiated at angles of less than 0.5 mrad 
with respect to the positron beam direction. 

The ZEUS apparatus is described more fully elsewhere~\cite{r1}.  Of particular importance 
in the present work are the central tracking detector (CTD), the vertex
detector (VXD) and the uranium-scintillator calorimeter (CAL). 

The CTD~\cite{CTD} is a cylindrical drift chamber situated 
inside a superconducting magnet coil which provides a 1.43 T field. 
It consists of 72 cylindrical layers covering the polar angle
region $15^{\circ} < \theta < 164^{\circ}$ and the radial  
range 18.2--79.4 cm.
The transverse momentum resolution for tracks traversing all CTD layers
is $\sigma(p_T)/p_T \approx \sqrt{(0.005 p_T)^2 + (0.016)^2}$, with $p_T$
in GeV. The VXD~\cite{VXD}  supplied tracking inside the CTD, 
and consisted of 120 radial cells, each with 12 sense wires covering the radial 
range 10.6--14.3 cm.  The vertex position of a typical multiparticle 
event is determined from the tracks
to an accuracy of typically $\pm$1 mm in the $X, Y$ plane and $\pm$4 mm in $Z$.

The CAL \cite{UCAL} gives an angular coverage of 99.7\% of $4\pi$ and  is divided
into three parts (FCAL, BCAL, RCAL), covering the forward (proton direction), 
central and rear polar angle ranges $2.6^{\circ}$--$36.7^{\circ}$,
$36.7^{\circ}$--$129.1^{\circ}$ and $129.1^{\circ}$--$176.2^{\circ}$, respectively. 
Each part consists of towers which are longitudinally subdivided
into electromagnetic (EMC) and hadronic (HAC) readout cells.
From test beam data, energy resolutions of
$\sigma_{E}/E = 0.18/\sqrt{E}$ for electrons and $\sigma_{E}/E = 0.35/\sqrt{E}$
for hadrons have been obtained (with $E$ in GeV). The calorimeter 
cells also provide time measurements which are used for beam-gas background 
rejection.


\section{Trigger and event selection} 

To identify jets in the event trigger and the subsequent selection of events, 
a cone algorithm~\cite{r5}  in accordance with the Snowmass
Convention~\cite{sno} was applied to the calorimeter cells. 
Each cell signal was treated as corresponding to a massless particle.
A cone radius of 1.0 in $R = \sqrt{(\delta\phi)^2 + (\delta\eta)^2}$ was used,
where $\delta\phi,\,\delta\eta$ denote the distances  of cells from the
centre of the jet in azimuth and pseudorapidity. 
A similar algorithm was used both online and offline.  
The transverse energy of the reconstructed jet is the  
sum of the measured transverse energies of the calorimeter cells included 
in it, and will be referred to as \eTrec.

The ZEUS detector uses a three-level trigger system.  The first level
trigger selected events on the basis of a coincidence of a regional or
transverse energy sum in the calorimeter and a track in the CTD pointing
towards the interaction point.  At the second level, at least 8 GeV of
transverse energy was demanded, excluding the eight calorimeter towers
surrounding the forward beam pipe.   Beam-gas background is further reduced
using the measured times of energy deposits and the summed energies in
the calorimeter.  At the third level, jets were identified.
The trigger combination used in the present study required 
at  least one jet with $\eTrec > 6.5$ GeV and $\eta<2.5$, or with 
 $\eTrec > 5.5$ GeV and $\eta<2.0$.
Cosmic ray events were rejected by means of information from the CTD   
 and the calorimeter.  An interaction vertex at a position $Z>-75$ cm,
as determined from the CTD tracks,  was demanded.  
A total of 392k events passed these requirements. 
The trigger efficiency is close to 100\%
over the entire kinematic range of events used in the present analysis. 

To study the association of \kzs\ mesons with jets, 
it was necessary to obtain a sample of high energy jets 
sufficiently centred in the acceptance of the CTD so as to 
optimise the acceptance for the $\pi^+$ and the $\pi^-$ decay products of 
associated \kzs\  mesons.  With these considerations in mind, 
events were selected offline with at least one jet 
having $\eTrec>7$ GeV and $|\eta|<0.5$. 
Standard ZEUS background rejection criteria were applied~\cite{r4,r1} to improve the 
rejection of beam-gas
events and cosmic ray events by means of cuts on the primary vertex position,
the fraction of well-measured tracks, the CAL signal times 
and the transverse momentum imbalance in the event.
On the assumption that the outgoing positron is not detected in the CAL, 
$y_{JB} = \sum(E- p_Z)/2E_e$ was calculated,  where the sum is 
over all calorimeter cells, treating each signal as equivalent to a massless 
particle; i.e.\ $E$ is the energy deposited in the cell, and
$p_Z$ is the value of $E\cos\theta$.  The quantity    
$y_{JB}$ is then a measure of $y^{true} = E_{\gamma,\,in}/2E_e$, where 
$E_{\gamma,\,in}$ is the energy of the incident virtual photon.
In the case that a DIS positron is present, a value of approximately unity 
is obtained. Events containing a DIS positron were rejected by requiring that
(i) no scattered beam positron be identified in the CAL, and 
(ii) $y_{JB} < 0.7$.
A requirement of $y_{JB}\ge 0.15$ was also imposed as part of the beam-gas 
background rejection procedure.  A total of 34.8k events was obtained at 
this stage, containing 36.6k jets with $\eTrec>7$ GeV and $|\eta|<0.5$. 
This selection of events, each containing at least one
accepted jet, represents the basic event sample for the analysis that follows. 

\section{Charged particle selection}
For the study of inclusive charged particle distributions, CTD tracks were accepted if
\begin{numlis}
\item they were associated with the primary vertex, 
\item their pseudorapidity was in the range $|\eta| \le 1.5$, and 
\item their transverse momentum satisfied $p_T\ge 0.5$ GeV. 
\end{numlis}
The pseudorapidity and momentum conditions are chosen to 
be the same as those to be imposed on the \kzs\ mesons. 
Apart from a small component arising from short-lived particle
decays, these conditions provide a sample of well-measured tracks which 
can be identified with charged hadrons originating 
from the main interaction.  In the following sections, charged particles 
will be referred to as $h^\pm$.

\section{$K^0$ reconstruction}

$K^0$ mesons decay 50\% as \kzs, which have a 68.6\% branching 
ratio into $\pi^+\pi^-$~\cite{PDG}.
\kzs\ mesons were identified by their charged decay mode $K^0_s\to
\pi^+\pi^-$. This decay mode is easily identifiable given the accurate 
tracking measurements from the CTD, since the mean decay
distance, projected on to the $r\phi$ plane, of $2.67p_T/m_K$ cm ensures
a spatial separation of the decay vertex from the primary event vertex
for a large number of the \kzs\ produced in the kinematic conditions of the 
present analysis. 

\kzs\ identification starts by selecting pairs of oppositely charged tracks 
obtained using the standard ZEUS reconstruction algorithms.
In order to restrict the track selection to the kinematic region where the tracking
was best understood, each track was required to satisfy 
the conditions $p_T> 150$ MeV and pseudorapidity $|\eta|<1.75$.
More details of the track reconstruction procedure are given in
ref~\cite{ZDISK}. Starting with the track
parameters determined at the point where the track passes nearest to the
beamline, and assuming the tracks in this region to be circular in 
the $r\phi$ plane,  candidate secondary vertices were then found 
by first calculating the intersection points in the $r\phi$ plane of all pairs 
of oppositely charged tracks in  an event.  Zero or two such intersection
points are found for a given track pair.  The $Z$-coordinates 
and the momentum components of each track of a pair were then calculated at 
each of these $(r,\phi)$ points.  

Further selections were made on the kinematic quantities listed below. 
%

\begin{romlis}
\item {\em Separation in $Z$ position.}  The separation $|\Delta Z|$ 
of the two tracks at a candidate secondary vertex was  
required to be less than 3 cm.   If a given track pair gave
two candidate secondary vertices, the one  with smaller $|\Delta Z|$ 
was chosen.
\item{\em Collinearity.}  The collinearity angle $\alpha$ is defined as the
projected angle in the $r\phi$ plane between the \kzs\ momentum and the 
line joining the primary to the secondary vertex.  Candidates were accepted 
if $\cos\alpha \ge 0.99$.
\item{\em Track impact parameter.} The impact parameter $\epsilon$ of a track
is defined as the distance of closest approach between the extrapolated 
track and the primary event vertex.  
A requirement $\epsilon > 0.3$ cm was imposed on both tracks for each 
kaon candidate.  
\item{\em Photon conversion.}  
The effective mass of the two tracks was evaluated assigning zero mass
to each track.  Track pairs whose effective mass was less than 50 MeV were
excluded from further consideration.
\item{\em $\Lambda$ removal.} The $p\pi$ mass hypothesis was applied to each
track pair, taking the higher momentum particle to be the $p$ ($\bar p$).
As in \protect\cite{ZDISK}, 
\kzs\ candidates with a $p\pi$ mass less than 1.12 GeV were rejected. 
\end{romlis}

\begin{figure}[t] 
\centerline{\epsfig{file=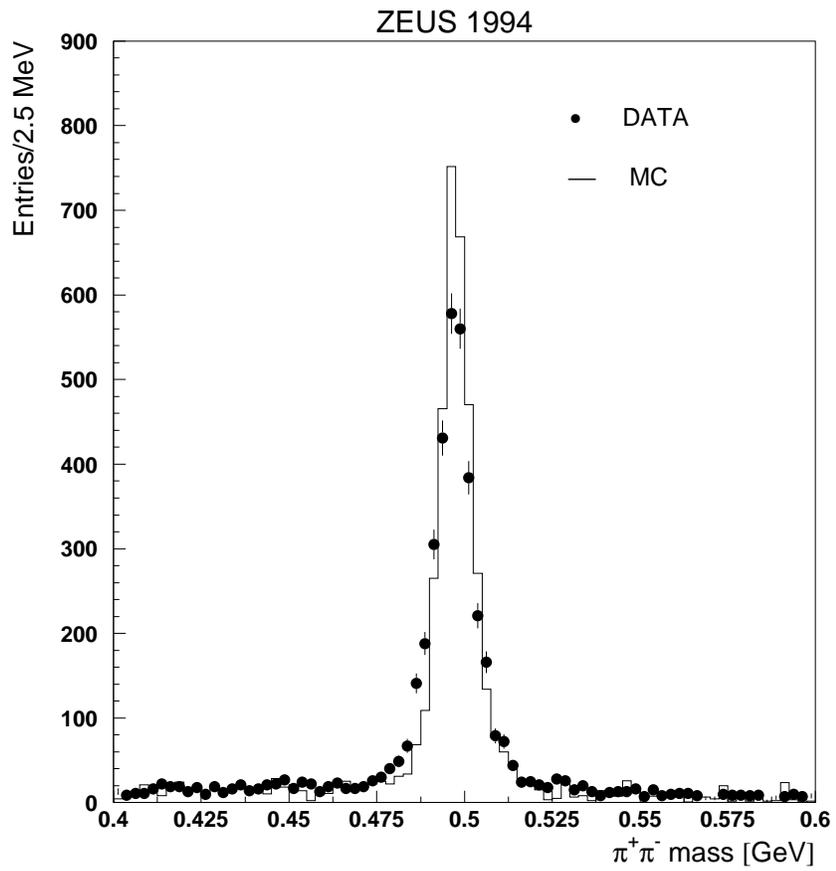,width=11.0cm,%
bbllx=40pt,bblly=190pt,bburx=500pt,bbury=670pt,clip=yes} }
\caption{\footnotesize
Reconstructed $\pi^+\pi^-$ mass distribution at selected secondary vertices,
compared with Monte Carlo prediction with equal normalisation.
\label{fc}}\end{figure}

\kzs\ mesons with $\pT \ge 0.5$ GeV  
and $|\eta|\le 1.5$ were used in the present analysis.  
The $\pi^+\pi^-$ mass distribution for all accepted candidates 
in this kinematic range is shown in fig.\ \ref{fc}, and displays a strong 
\kzs\ signal. After subtracting a background (averaged over windows
between 440--470 and 530--560 MeV) from the signal region of 470--530 MeV,
a total of 3154$\pm$63 \kzs\ was evaluated on a background of 388.  
The mean value of the reconstructed \kzs\ mass was 497.0$\pm$0.1 MeV 
(statistical error) compared with the
nominal value of 497.7 MeV~\cite{PDG}, with a fitted width 
6.3$\pm$0.1 MeV.   Monte Carlo comparisons show that the decay vertex 
and momentum of the {\kzs} are well reconstructed over the full range 
of momentum and pseudorapidity, without significant
systematic bias.  This background subtraction method was used to obtain
the  \kzs\ signal in each bin of all plotted distributions.
Further details of the \kzs\ reconstruction may be found in ref.~\cite{theses}.

\section{Monte Carlo simulations and data correction}

The experimental data were compared with Monte Carlo 
generated events obtained using PYTHIA 5.7, with JETSET 7.4 for the 
hadronisation~\cite{Pythia}.  Systematic studies were  
made using HERWIG 5.8~\cite{Herwig}.   PYTHIA is found to give a reasonable 
description of jet profiles in hard photoproduction in the 
kinematic region of the present study~\cite{r6, r7}. PYTHIA events were
generated using a $p_{T\;min}$ value of 2.5 GeV as the cut-off for the 
hard subprocess; the results were insensitive to this choice. 
The event generation was followed by a full simulation of the 
detector and trigger response in ZEUS by means of GEANT 3.13~\cite{GEANT}.  
The direct and resolved contributions are combined in proportion to their 
generated cross sections.
The GRV-LHO parton densities for the photon~\cite{GRV} and MRSA for the 
proton~\cite{MRS} were used in running the standard version of 
PYTHIA.  Variations on the standard calculation were made by reweighting the
results to use different photon parton densities, 
and also by using an option which uses 
multiparton interactions (MI) as a model for an underlying event 
accompanying the main hard QCD subprocess (with a $p_{T\;min}$ value 
of 1.4 GeV for the secondary interactions).  In the LO QCD model employed here, 
multiparton scattering may accompany the resolved
processes, in which the photon acts as a source of partons, 
 but cannot accompany the direct process, in which the photon interacts as 
a pointlike entity with no partonic substructure.  

Within string fragmentation models such as PYTHIA, the most important quantity governing 
the number of strange particles that appear in the fragmentation is the
so-called ``strange\-ness suppression parameter" $P_s/P_u$. This is  
assigned a default value of 0.3 on the basis of measurements at PETRA and
PEP~\cite{Saxon}, which measured a variety of strange/non-strange 
particle ratios (e.g.\ $K^0$ to $\pi^\pm$ mesons)
in the products of $e^+e^-$ annihilation.  
In previous studies of neutral kaons in DIS at HERA~\cite{ZDISK, HDISK} it
was found that decreasing $P_s/P_u$ from its 
standard value improved the agreement between the Monte Carlo and the data. 
We have therefore generated Monte Carlo event samples also 
with the value of $P_s/P_u$ decreased from 0.3 to 0.2.   

The Monte Carlo events were used to determine correction factors
for the data.  These were calculated separately for each bin  of
any given plot, so that for each plotted quantity the corrections are suitably
averaged over the other physical quantities.  For \kz\ 
calculations, the following factors were taken into account:
\begin{romlis}
\item {\em \kzs\ reconstruction efficiency $\epsilon_{K^0_s}$.}  
Given a set of reconstructed and accepted Monte Carlo events, this
was the ratio of (a) the number of    
reconstructed $K^0_s\to\pi^+\pi^-$ entries in a given bin, background 
subtracted, to (b) the corresponding number of  
generated $K^0_s\to\pi^+\pi^-$ occurring in this bin.    
All the other event selection criteria are applied in the normal way.
\item{\em Reconstruction level/hadron level correction factor $C$.}  
This factor makes use of jets reconstructed from the four-vectors of the 
primary final state particles (charged or uncharged) in the generated  
events, referred to as ``hadron jets".  For events with a
generated $K^0_s\to\pi^+\pi^-$ in a given bin,   
$C$ is defined as the ratio of (a) the number of events 
which satisfy the trigger and acceptance conditions and  
have a reconstructed calorimeter jet with  $\eTrec>7$ GeV and $|\eta|<0.5$, 
to (b) the number of events having $0.2 \le y^{true} < 0.85$  and a hadron jet 
with cone radius 1.0, $\eT\ge 8$ GeV and $|\eta| \le 0.5$.
In this way we correct to a defined set of kinematic conditions 
at the final state level. The parameters in (b) are chosen
to provide a good correspondence with the cuts at the detector level, and
thereby minimise the corrections.  The $y^{true}$ range corresponds to a  
$\gamma p$ centre of mass energy range of 134--277 GeV.
\end{romlis} 

Since the shapes of the distributions
are similar in the data and Monte Carlo (after reconstruction), 
the calculated correction factors will suitably  
take into account all effects due to reconstruction efficiency, 
event selection efficiency and bin-to-bin migration.  These aspects have 
been checked in our previous studies of inclusive jet production~\cite{r5}. 
The \kzs\ reconstruction efficiency is 
dominated by the geometric effects of the track cuts and 
the CTD tracking efficiency, with little sensitivity to the 
properties of the jet.  It is similar for direct and resolved events. 
In the region of $p_T$ and $\eta$ used here, 
$\epsilon_{K^0_s}$ has a plateau of approximately 0.35  in the middle of the 
accepted range of either quantity, falling to 0.25 at the ends.  
$C$ takes values typically between 0.6 and 0.8.   

Corrected numbers of $K^0$ mesons in bins $\Delta v$ of any 
given variable $v$ are evaluated according to the formula
\begin{equation}  
\frac{dN}{dv}\!\!
\begin{array}{l}\mathit{(corrected)}\\ \, \end{array}  = 
\frac{N\mathit{(detected)}}{\Delta v\,\epsilon_{tot}\,B}, 
\end{equation}
where $\epsilon_{tot}=\epsilon_{K^0_s}\,C$, and $B$ is the 
total branching ratio $K^0\to\kzs\to \pi^+\pi^-$.

For charged particles, a similar relationship 
$\epsilon_{tot}=\epsilon_{h^\pm}\,C$ is used, 
where the charged particle reconstruction efficiency
$\epsilon_{h^\pm}$ is calculated  as the ratio of reconstructed and 
accepted tracks to the number of generated charged particles in the same 
bin.  $C$ is then calculated analogously to the \kz\ case, and $B$ is set to
unity.  Here and in the calculation of $C$, generated charged particles 
are taken to be those in the final state with lifetimes greater than
$10^{-8}$ s, together with charged decay products of primary 
hadrons with a shorter lifetime than this, 
but excluding the products of \kzs\ and $\Lambda,\,\bar\Lambda$ decays. 

All the particle distributions are normalised to the number of jets. 
Keeping in mind that each event in the basic event sample has at least 
one accepted jet, we evaluate (a) the total number of 
charged  particles or kaons in each bin of a given distribution, 
and (b) the total number of accepted jets in the basic event sample.
The ratio of (a) to (b) is then plotted to give numbers of particles per jet. 
To obtain corrected distributions, it is thus 
necessary to correct the total number of measured jets 
$N_{jets}$, in the defined kinematic range, 
to the total number of above-defined hadron jets. To achieve this,  
the correction formula (1) is again employed, using an efficiency factor                            
$\epsilon_{tot}= C$ together with $B=1$, and removing from 
the definition of $C$ the conditions on kaons or charged particles.

\section{Systematic uncertainties}

The stability of the results against reasonable variations of the selection
criteria was studied in order to estimate the systematic uncertainties. 
The most important effects were estimated as follows:
\begin{numlis}
\item To study the sensitivity to the fragmentation scheme in the Monte Carlo, 
HERWIG was used instead of PYTHIA to evaluate the correction factors.
\item An uncertainty exists in the matching of the calorimeter 
energy scale in the Monte Carlo events to that of the data.  To estimate this, 
the Monte Carlo energy scale was varied by $\pm 5$\%.  
\item To estimate tracking uncertainties, the effect  of including or not 
including the VXD in the track reconstruction was investigated.
\item The effect of varying the accepted
$y_{JB}$ range to $0.2 < y_{JB} < 0.8$ was evaluated.
\item{$K^0$ definition:}\\ 
(i) the cut on $|\Delta Z|$ was varied
between 2 and 4 cm;\\ 
(ii) the cut on $\cos\alpha$ was varied by $\pm 0.005$;\\
(iii) the cut on the track impact parameter $|\epsilon|$
was varied in the range 0.27--0.33 cm;\\ 
(iv) the upper mass value for $\Lambda$ rejection was varied between 1.117 and
1.123 GeV.
\end{numlis}

The effects on the results of most of the parameter variations listed
were normally found to be small, at the level of up to a few percent.
The individual contributions were combined in quadrature to give the
total systematic error on a given point.
The systematic errors on the correction factors are 
dominated by the effects of (1), (2) and (3).
The comparison of uncorrected data with PYTHIA predictions is affected 
by  (2) and (3).

\section{Results}
\subsection{Charged particles}

\begin{figure}[ht]
\centerline{\epsfig{file=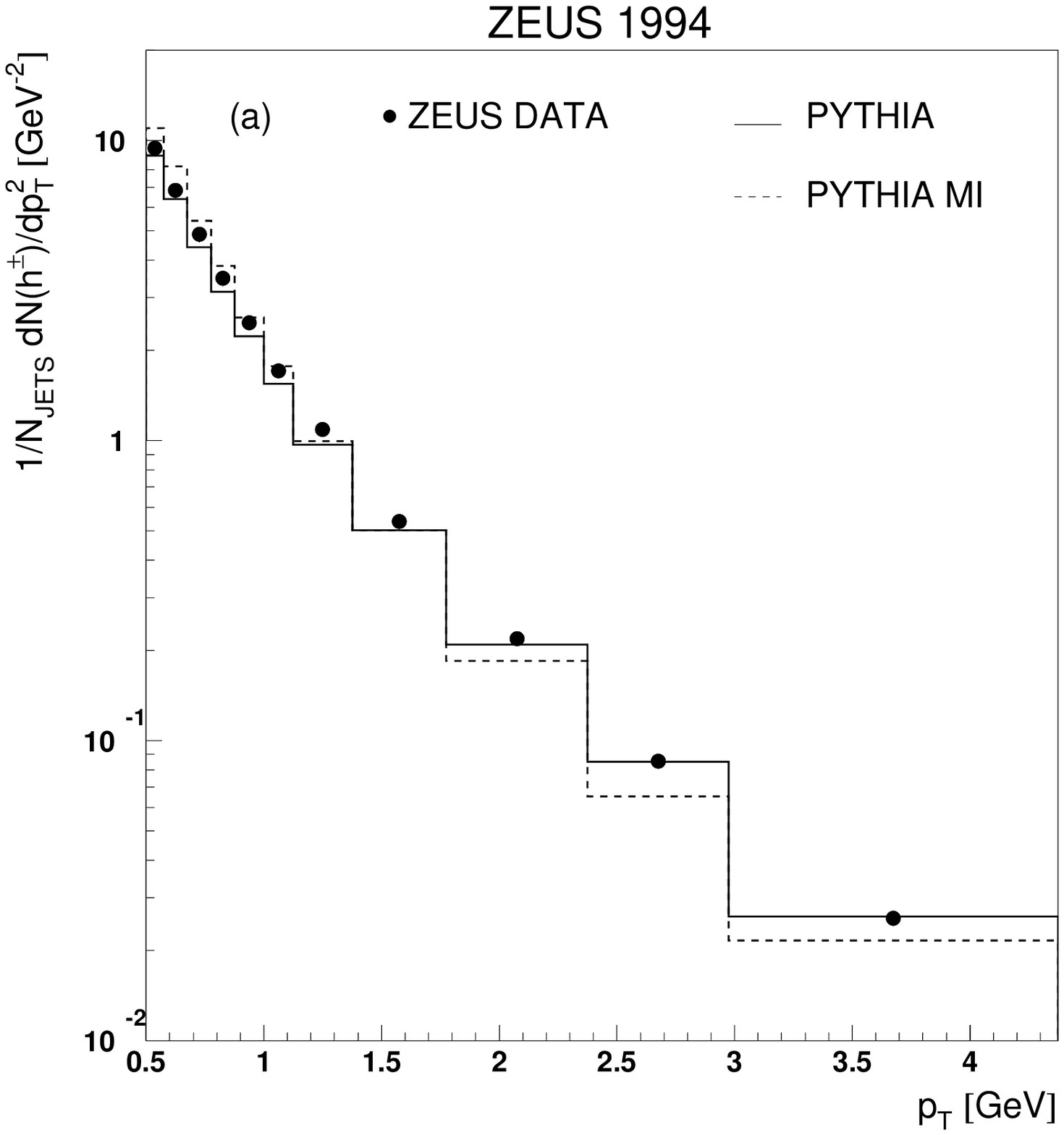,width=10.3cm,%
bbllx=35pt,bblly=190pt,bburx=500pt,bbury=675pt,clip=yes} }
\centerline{\epsfig{file=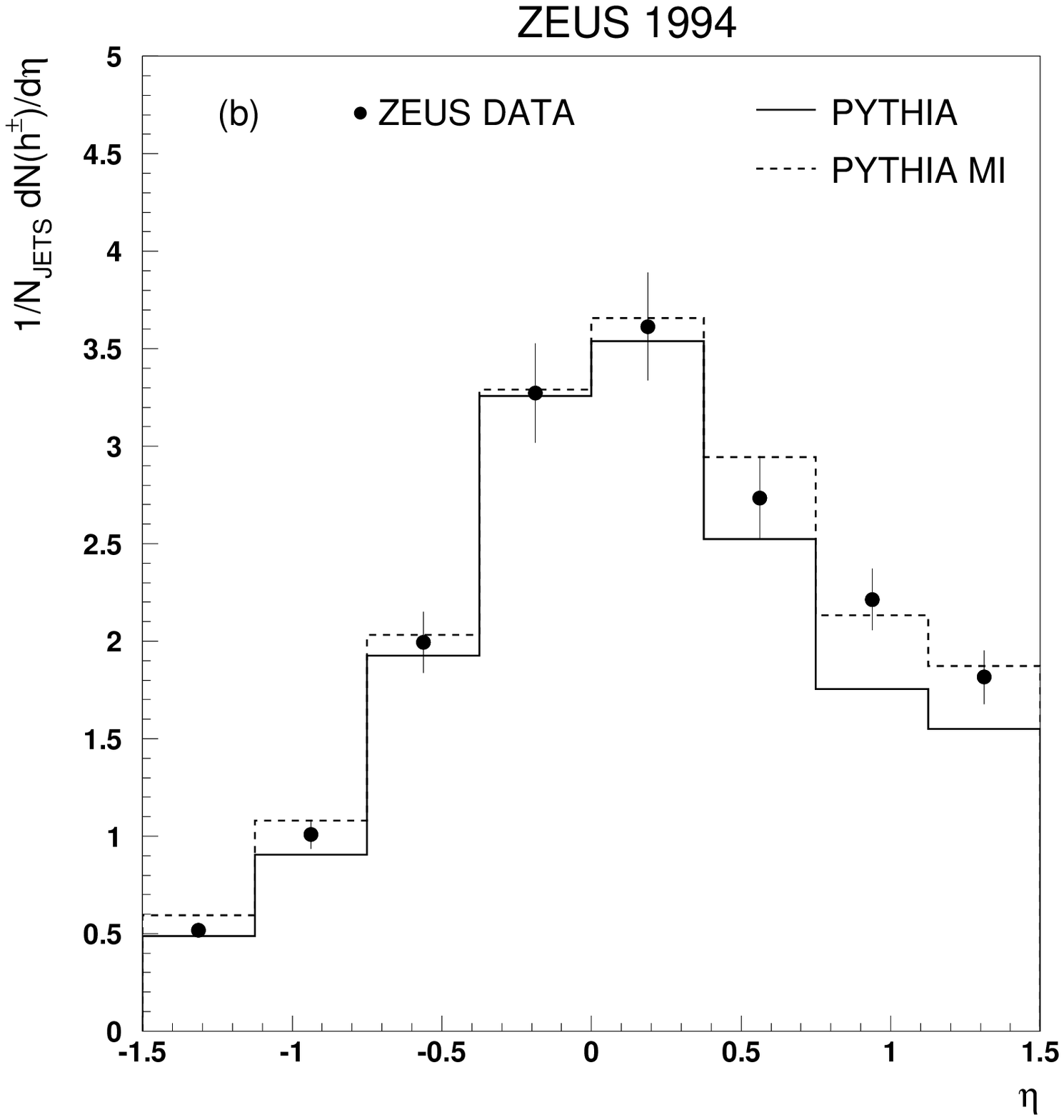,width=10.3cm,%
bbllx=35pt,bblly=190pt,bburx=500pt,bbury=675pt,clip=yes} }
\caption{\footnotesize Corrected distributions of charged particles 
per jet in events containing a hadron jet with $E_T\mbox{(jet)}\ge 8$ GeV 
and $|\etaj|\le 0.5$. (a) $(1/N_{jets})\,dN(h^\pm)/dp_T^{\,2}$ 
as a function of $\pTh$ for $|\etah|\le 1.5$; (b) as a function of $\etah$ for $\pTh\ge 0.5$ GeV. 
Error bars are statistical and systematic combined in quadrature, and are dominated 
by the systematic.   The PYTHIA MI Monte Carlo uses a multiparton
interaction model.
\label{fi}}\end{figure}

In fig.\ \ref{fi} we present efficiency and acceptance  
corrected distributions of $dN(h^\pm)/dp_T^{\; 2}$ and $dN(h^\pm)/d\eta$ 
in $\pTh$ and $\etah$ respectively.  As defined in section 6, 
the corrected distributions are given for particles in events 
containing a final state hadron jet of $\eTj \ge 8 $ GeV 
with $|\etaj|\le 0.5$.  Here and throughout, all results  
are normalised as  numbers of particles per jet satisfying the
stated definitions. 
The $\pTh$ distributions are averaged over $ -1.5 \le \etah < 1.5$;
the $\etah$ distributions are averaged over $\pTh \ge 0.5$ GeV. 
At this stage no explicit association of the $h^\pm$ with the jets is made.
The statistical errors are small. 

Good agreement is seen between the data and the standard PYTHIA 
predictions in the shape and magnitude of the \pT\ distribution.
In the $\eta$ distribution, the overall agreement is still good, but
it is apparent that standard PYTHIA tends to undershoot the data at 
higher $\eta$ values.  A similar trend has been observed in
other photoproduction studies at HERA~\cite{H1MI,r5,r6}, 
leading to suggestions that the discrepancy can be attributed 
at least in part to multiparton interactions~\cite{ZEUSMI}.  The MI option 
is seen to give an improved description of the data at high $\eta$; however, 
it overestimates the exponential slope of the  \pT\ distribution.
A reduction in the  $P_s/P_u$ value has a negligible 
effect on the present distributions (not shown). Little difference is seen 
if different photon parton densities are used. 

\begin{figure}[ht]
\centerline{\epsfig{file=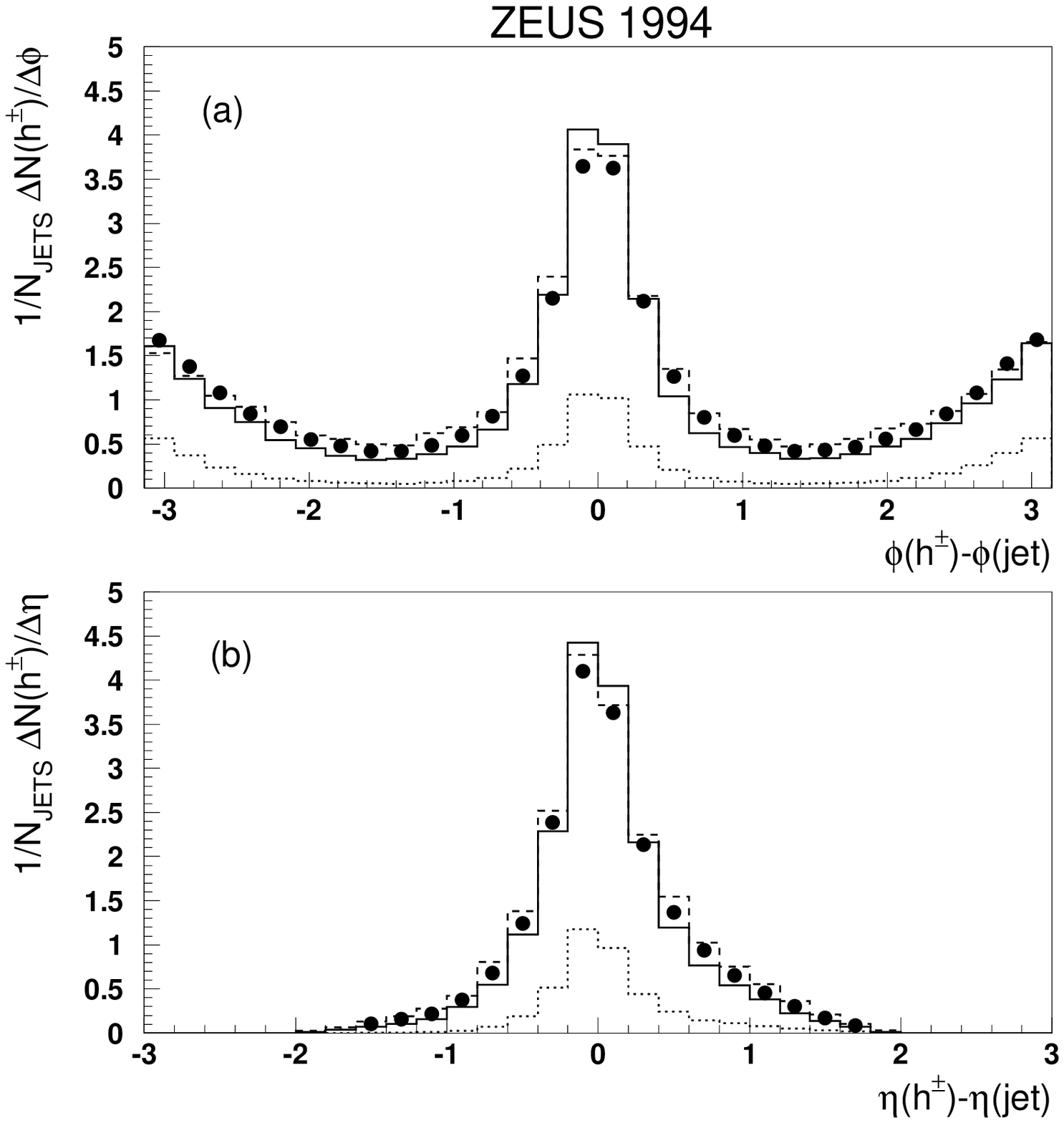,width=10.3cm,%
bbllx=35pt,bblly=190pt,bburx=500pt,bbury=675pt,clip=yes} }
\centerline{\epsfig{file=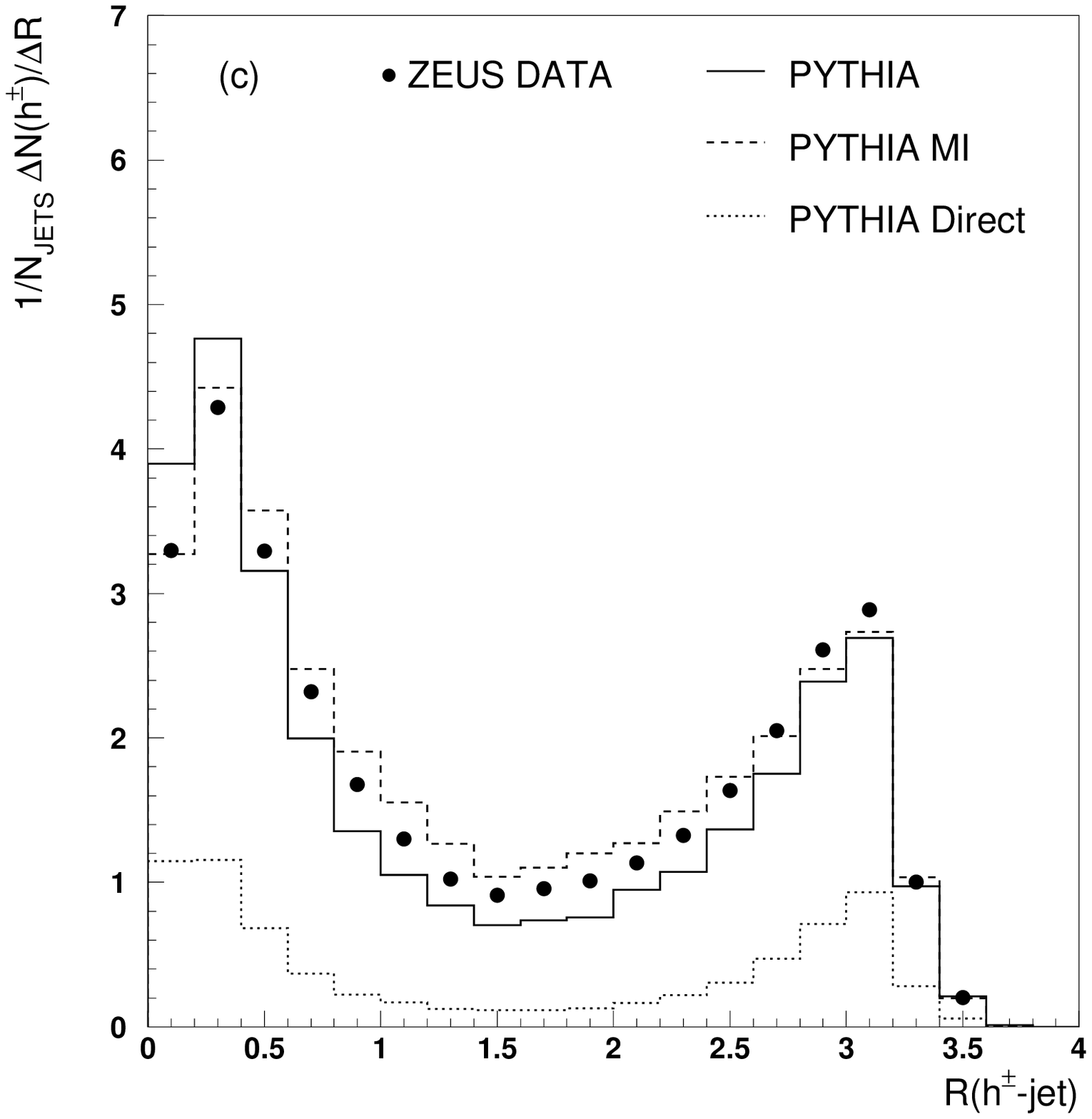,width=10.3cm,%
bbllx=35pt,bblly=190pt,bburx=500pt,bbury=660pt,clip=yes} }
\caption{\footnotesize Distance of charged particles (h$^\pm$) from jet (a) 
in $\phi$, (b) in $\eta$, for $|\Delta\phi| < \pi/2$, (c) in $(\eta,\phi)$,    
with comparison to various Monte Carlo predictions as indicated.
The entries are uncorrected numbers of charged particles ($p_T\ge 0.5$ GeV, 
$|\eta|\le 1.5$) per measured jet ($\eTrec\ge 7$ GeV, 
$|\eta|\le 0.5$) per unit interval of the plotted quantity. 
The histograms correspond to PYTHIA predictions 
using the GRV-LHO photon structure and default strangeness suppression. 
The direct  contribution to the standard PYTHIA calculation 
is also shown separately.
A common systematic error of 5\% should be allowed on the data points
in comparing with the Monte Carlo.
\label{fj}}
\end{figure}

We now examine in more detail the association of 
charged particles with jets, given the presence of at least one jet in
each event in the selected data sample.  
Fig.\ \ref{fj}(a) shows the uncorrected distribution in the difference 
$\Delta\phi$ in azimuth between an $h^\pm$ and the axis of 
any jet in the event satisfying $\eTrec \ge 7$ GeV 
and $|\etaj|\le 0.5$. The peak around $|\Delta\phi| =\pi$ indicates the 
presence of jets opposite in azimuth to the observed particle. 
For $h^\pm$-jet pairs with $|\Delta\phi| < \pi/2$, the distance $\Delta\eta$ 
in pseudorapidity between the $h^\pm$ and the jet is plotted in 
fig.\ \ref{fj}(b). In comparing the data to the Monte Carlo predictions, 
an overall systematic uncertainty of $\pm5\%$ should be allowed.

In discriminating between the various models, a particularly useful 
variable was found to be $R = \sqrt{(\Delta\phi)^2 +(\Delta\eta)^2}$, 
i.e.\  the distance in $(\eta, \phi)$ between a given particle 
and the jet axis.  The uncorrected $R$ distribution is presented in 
fig.\ \ref{fj}(c). A prominent peak at small $R$ demonstrates that the $h^\pm$ 
are being produced in association with jets; there
is also a peak at $R\approx 3$ due to the likely existence of another jet 
opposite in $\phi$ to the first.  In the following discussion 
we concentrate mainly on the region $R < 2.5$. 
Over the whole range, the main contributions come from resolved processes.
The predictions of standard PYTHIA lie slightly above the data at low $R$, but 
are low in the inter-jet region $1 < R < 2.5$.
To establish whether the latter discrepancy is associated with 
the excess of \hpm\ over expectations seen at high \etah\, 
a similar plot was made with the \hpm\ acceptance changed to 
$-1.5 < \etah < 0.5$  (not shown).  This reduced the 
discrepancy for $1 < R < 2.5$ slightly but did not eliminate it.
Variations of the $P_s/P_u$ parameter and the photon
parton density again have little effect on the 
distributions.  The effect of using the MI option 
of PYTHIA is more pronounced; in the inter-jet region, the MI option 
overcompensates for the previous shortfall of standard PYTHIA events, 
while at small $R$ the jet appears broadened.  
Overall, the MI option provides an improved description of the data.

The mean corrected multiplicity of charged particles with $\pTh \ge 0.5$ GeV 
per jet, integrated over $R\le 1$, 
is shown in Table~\ref{tbl} compared with different results from PYTHIA.
All the models are consistent with the experimental result, within errors. 

\subsection{\kz\ mesons} 

\begin{figure}[ht] 
\centerline{\epsfig{file=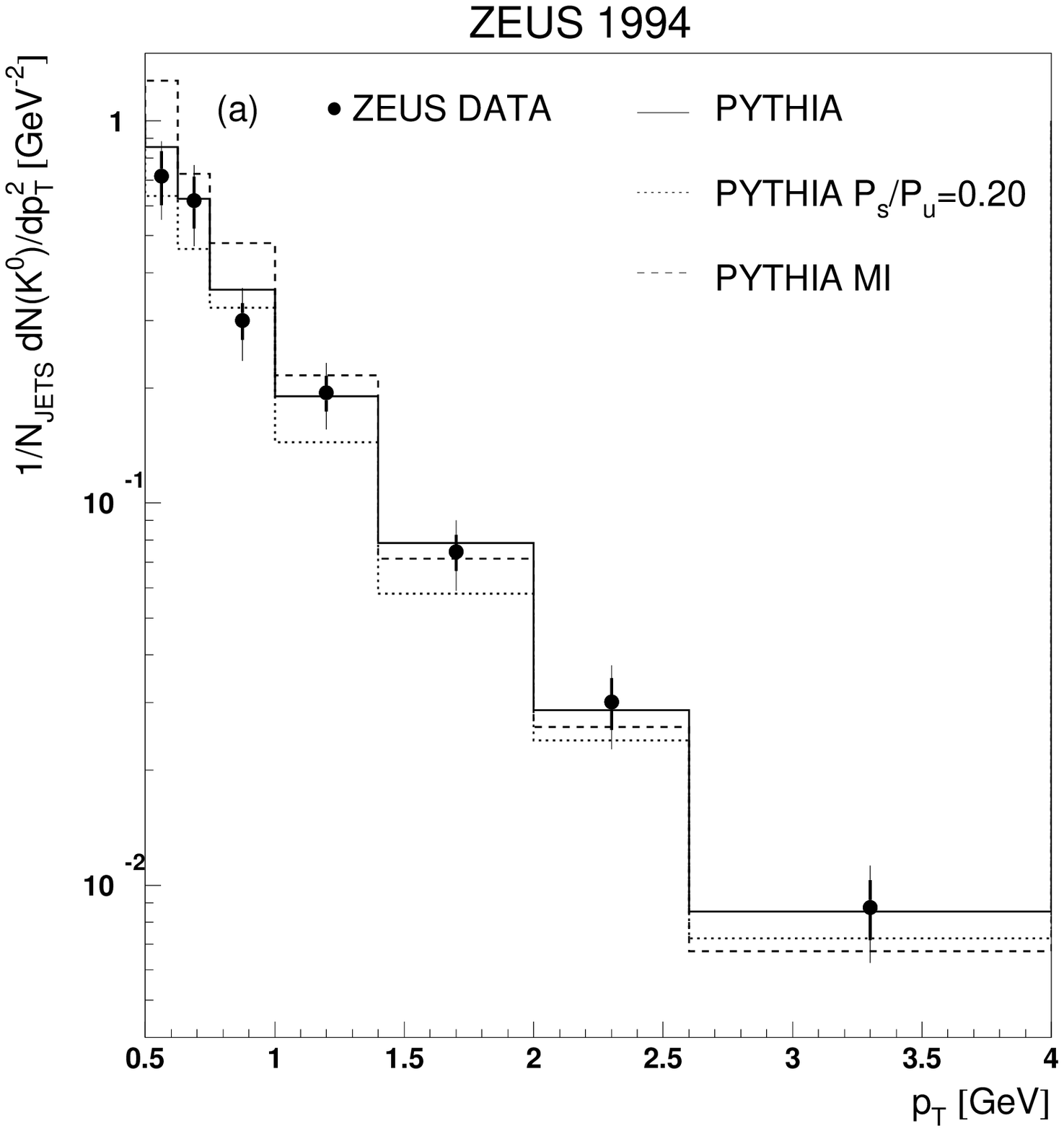,width=10.3cm,%
bbllx=35pt,bblly=190pt,bburx=500pt,bbury=675pt,clip=yes} }
\centerline{\epsfig{file=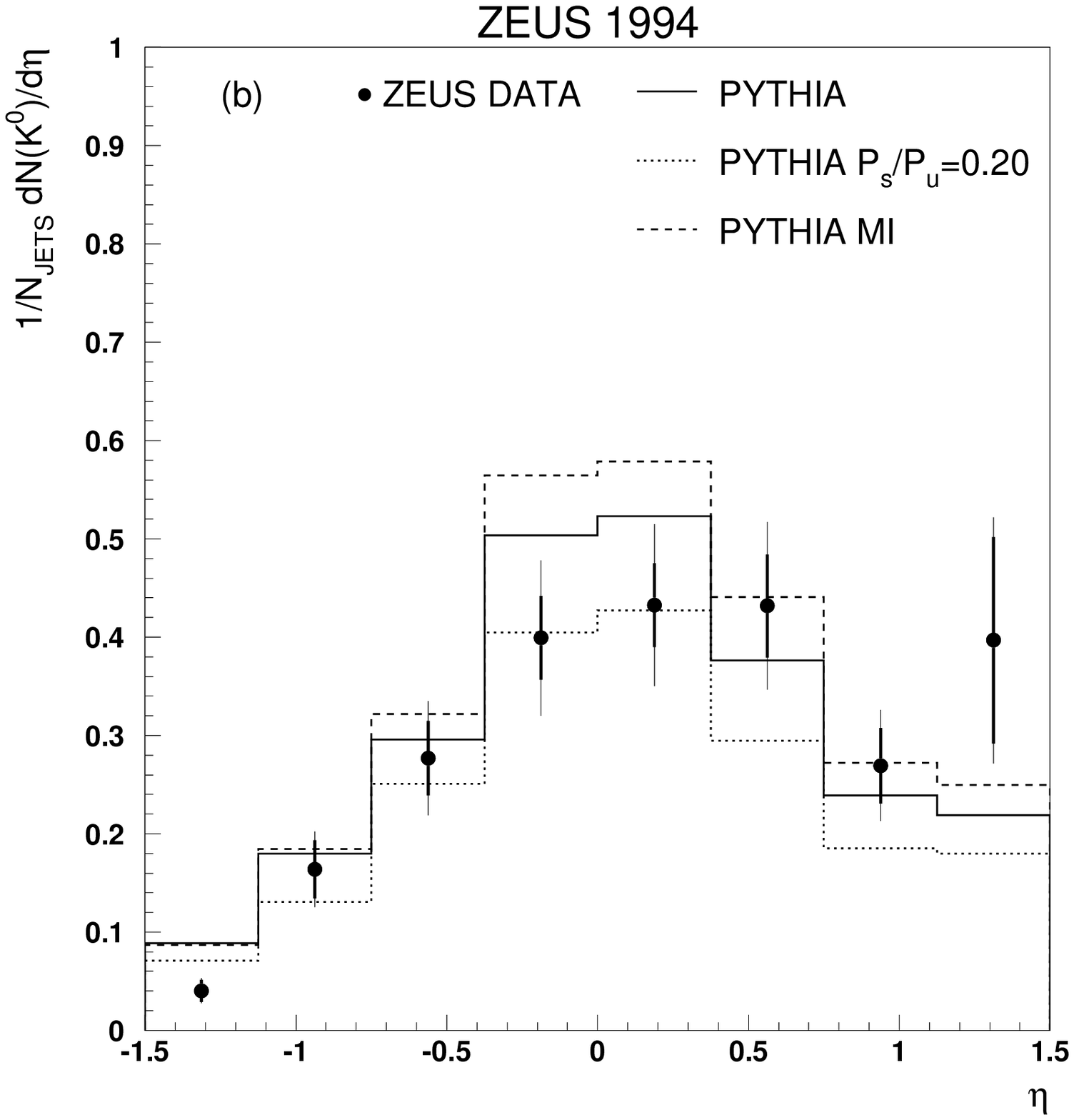,width=10.3cm,%
bbllx=35pt,bblly=190pt,bburx=500pt,bbury=675pt,clip=yes} }
\caption{\footnotesize Corrected distributions of $K^0$  mesons per jet in 
events containing a hadron jet with $\eTj\ge 8$ GeV 
and $|\etaj|\le 0.5$.
(a) $(1/N_{jets})\,dN(K^0)/dp_T^{\,2}$ as a function of $\pTk$ for $|\etak|\le 1.5$; 
(b) $(1/N_{jets})\,dN(K^0)/d\eta$ as a function of $\etak$ for $\pTk\ge 0.5$ GeV. 
Inner and outer error bars are statistical errors and statistical
combined with systematic in quadrature.
\label{fe}}\end{figure}

In the previous section it was seen that the features of charged particles 
in photoproduced jets are generally well described using standard PYTHIA, 
with some further improvement obtainable using the MI option.  Turning now 
to \kz\ mesons, we consider the same distributions as for the $h^\pm$.
As before, at least one reconstructed calorimeter
jet with $\eTrec\ge 7$ GeV and $|\etaj|\le0.5$ was demanded in an event, 
and the corrected distributions are for events containing a final-state 
hadron jet with $\eTj \ge 8 $ GeV and  $|\etaj|\le 0.5$.
The distributions are normalised as numbers of \kzs\ or \kz\ per jet
for uncorrected or corrected data respectively.  
Corrected $dN(K^0)/dp_T^{\; 2}$ and $dN(K^0)/d\eta$ distributions are
shown in fig.\ \ref{fe} as functions of $\pTk$ and $\etak$.
The $\pTk$ distributions are averaged over $-1.5 < \etak < 1.5 $; 
the $\etak$  distributions are averaged over $\pTk \ge 0.5$ GeV. 
No attempt is yet made to associate the \kzs\ with jets. 
In the larger error bars the systematic and statistical errors 
are summed in quadrature.  
Following the studies mentioned~\cite{ZDISK,HDISK}, we also show results 
with the value of $P_s/P_u$ decreased from 0.3 to 0.2.  
 
A reasonable overall agreement is seen between the data and the 
various PYTHIA plots in the shape and magnitude of the \pT\ spectrum,
although the MI option tends to overestimate the slope of the plot.
For $\etak\le 0.5$, the data of fig.\ \ref{fe}(b) show an approximate agreement 
with the standard PYTHIA which is improved by using a reduced value of $P_s/P_u$.   
At higher \etak, however, a tendency for the PYTHIA values to be low 
compared with the data is worsened when $P_s/P_u$ is reduced.
The agreement in this region is improved with the use of the MI option.

\begin{figure}[ht] 
\centerline{\epsfig{file=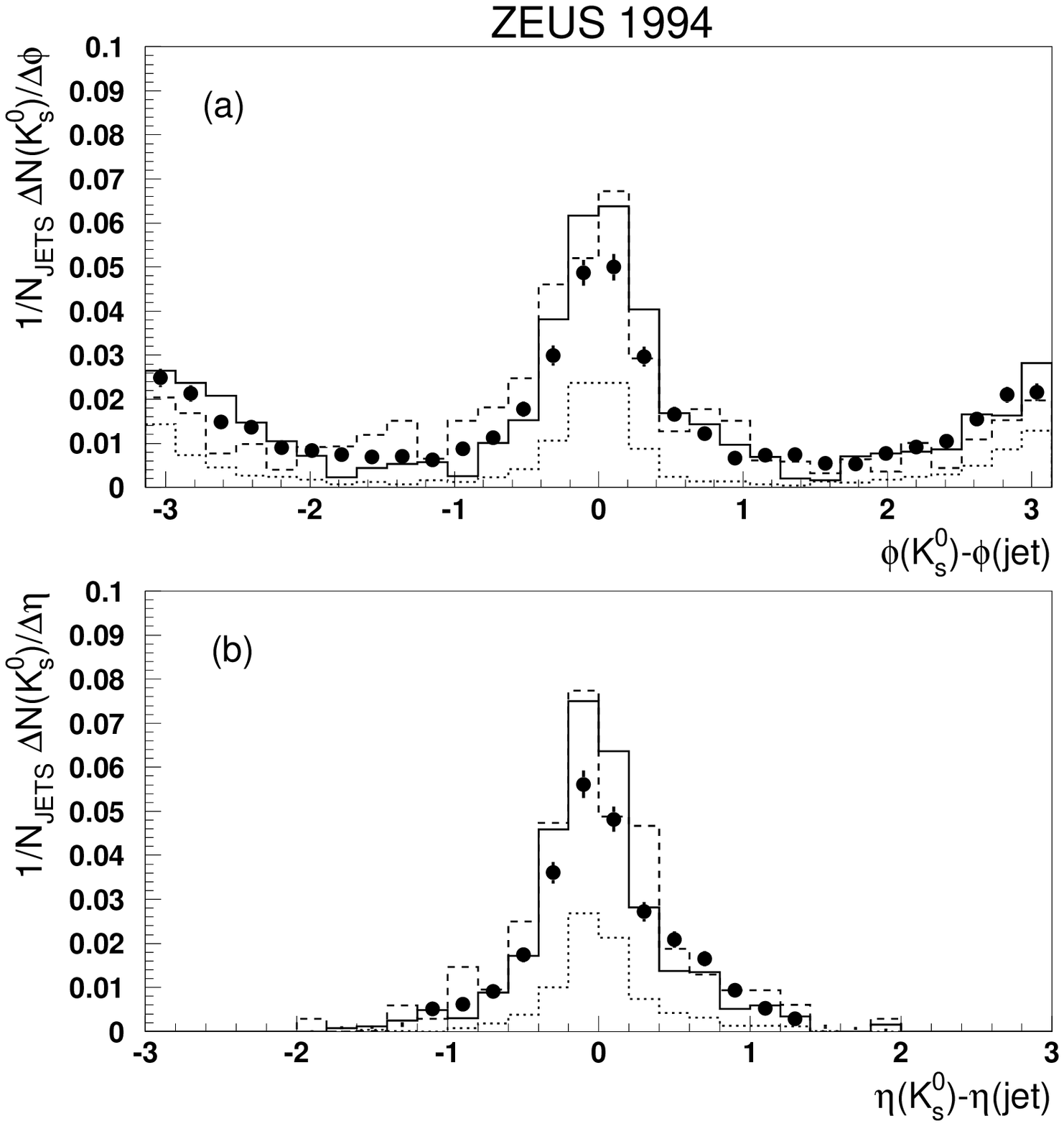,width=10.3cm,%
bbllx=35pt,bblly=190pt,bburx=500pt,bbury=675pt,clip=yes} }
\centerline{\epsfig{file=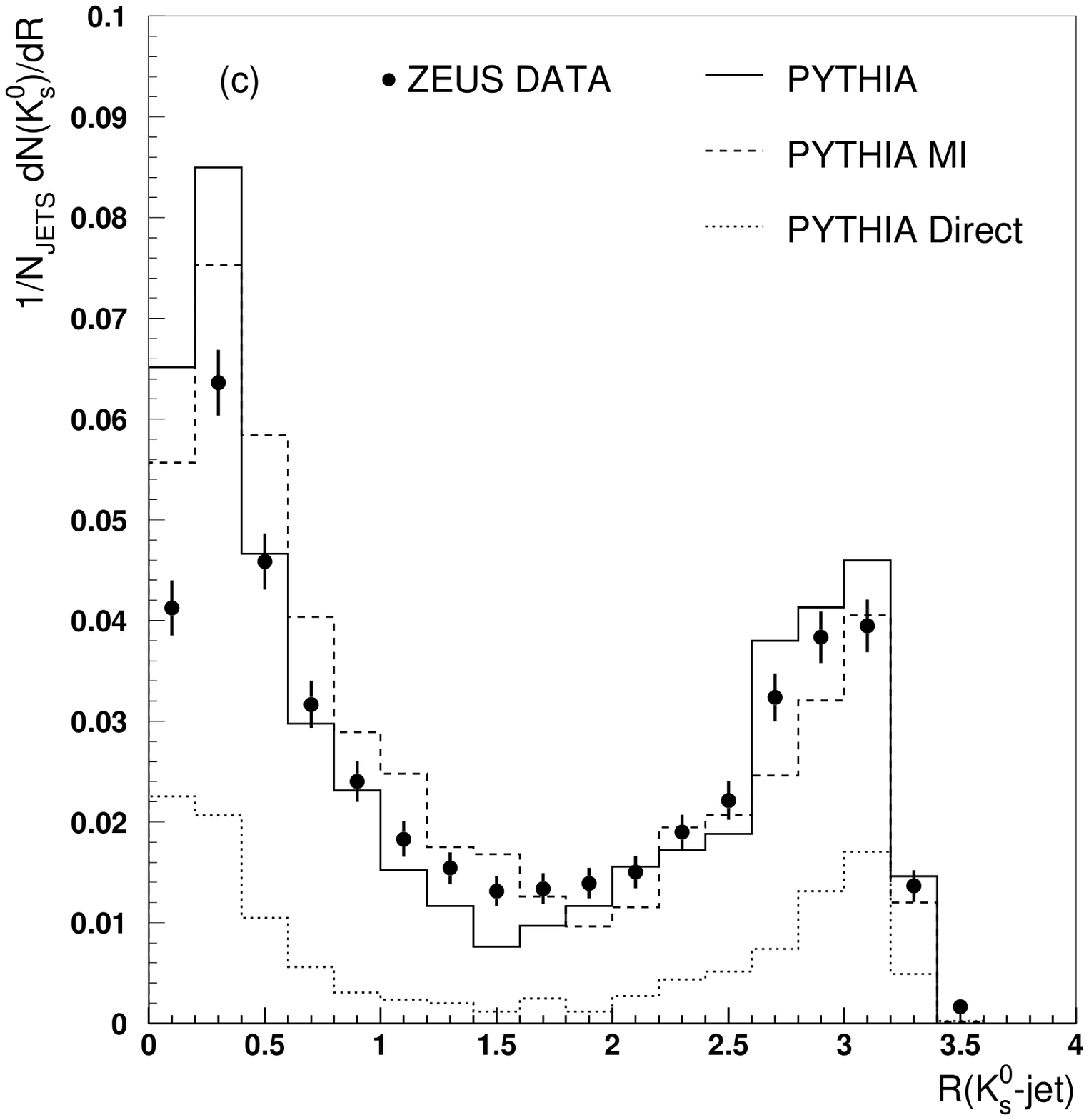,width=10.3cm,%
bbllx=35pt,bblly=190pt,bburx=500pt,bbury=660pt,clip=yes} }
\caption{\footnotesize Distance of $K^0_s$ from jet (a) 
in $\phi$, (b) in $\eta$, for $|\Delta\phi| < \pi/2$, (c) in $(\eta,\phi)$,    
with comparison to various Monte Carlo predictions as indicated.
The entries are uncorrected numbers of 
$K^0_s$  ($p_T\ge 0.5$ GeV, $|\eta|\le 1.5.$) 
per measured jet ($\eTrec\ge 7$ GeV, 
$|\eta|\le 0.5$) per unit interval of the plotted quantity. 
The histograms correspond to PYTHIA predictions 
using the GRV-LHO photon structure and default strangeness suppression. 
The direct  contribution to the standard PYTHIA calculation 
is also shown separately.
A common systematic error of 7\% should be allowed on the data points
in comparing with the Monte Carlo.
\label{ff}} \end{figure}

Fig.\ \ref{ff}(a) shows the uncorrected distribution in the difference 
$\Delta\phi$ in azimuth between a \kzs\ and
any jet in the event satisfying $\eTrec  > 7$ GeV  and $|\etaj|\le 0.5$.  
For \kzs-jet pairs with $|\Delta\phi| < \pi/2$, the distance $\Delta\eta$ 
in pseudorapidity between the \kzs\ and the jet is plotted in 
fig.\ \ref{ff}(b).     
A common overall systematic uncertainty of $\pm7\%$ should be allowed.
The data are reasonably well fitted by the Monte 
Carlo outside the peak at $\Delta\phi\approx 0$ and in the wings of the 
\etak\ distribution.
In both distributions, standard PYTHIA 
overestimates the numbers of kaons near the axis of a jet. 

The uncorrected distribution in $R$ is plotted in fig.\ \ref{ff}(c) 
for all \kzs.  
Here it is clear that standard PYTHIA overestimates the data in the jet core, 
i.e.\ for $R<0.4$, while underestimating it in the interjet region.  
%
To gain understanding of the possible reasons for the discrepancy 
in the jet core, a variety of investigations were made.  
It was found that the discrepancy persists
when the number of charged tracks in the region of the jet (i.e.\ with $R<1$)
is selected to be small: even with just one or two charged particles
present in addition to the \kzs, a similar effect is seen.  
In the events used in this analysis, the
mean number of additional charged particles in the jet is approximately 3.  It is
therefore difficult to attribute the effect to a poorly understood
problem with the charged particle tracking.  

The effects of varying the photon parton densities were investigated and found
to be small: the use of the   GRV-LO~\cite{GRV}, 
ACFGP~\cite{Aur}, LAC1~\cite{LAC} or GS-HO~\cite{PD}  parton sets 
altered the predictions by less than 5\% for $R < 1$.  
No attempt was made to vary the parton densities in the proton, 
since these are well defined in the present kinematic range from other 
measurements.  Removing the small contribution from
charm-containing jets in the generated events made only a slight 
difference, giving little scope for remedying the problem by remodelling the
charm simulation.  If the normalisation is performed to luminosity
rather than to numbers of jets, the conclusions are likewise unchanged. 
A similar although less marked discrepancy at low $R$ was found using
predictions from the HERWIG Monte Carlo.  
%
The discrepancy outside the jet is again found to be reduced 
slightly but not eliminated if the acceptance is reduced to 
$-1.5 < \etaks < 0.5$.   The MI option gives an improved description in 
the inter-jet region, but is worse elsewhere.  

In about 7\% of the kaonic events, two \kzs\ were found.
The distance $R$ in $(\eta,\phi)$ between them was 
plotted, giving a distribution which was 
fairly flat up to $R\approx 3.5$  with a small enhancement  at low $R$.
Standard PYTHIA tends to overestimate the data by approximately the
same amount as at $R< 0.4$ in fig. \ref{ff}(c), but over the entire
$R$ range.  The conclusion is that many of the \kz\ pairs are not strongly
correlated.   

The characteristics of the \kzs\ in jets can be further investigated 
by distinguishing between direct and resolved 
photoproduction processes, as defined at leading order in QCD. 
A subsample of events was chosen in which at least two jets were found,
one satisfying the above experimental jet definition and a  
second having $\eTrec > 7$ GeV as before, and $\etaj < 2.5$. The two 
highest \eT\ jets satisfying these conditions were used to estimate the 
fraction of the photon energy which takes part in the hard subprocess.
This estimate is given in terms of the measured parameters of the two jets as 
$\xgO=(E_{T\,1}^{\;rec}e^{-\eta_1} +E_{T\,2}^{\;rec}e^{-\eta_2})/2E_e\,y_{JB}$. 
For $\xgO < 0.75$ the event samples are 
dominated by the resolved process, above this value the direct process 
is more important~\cite{r6}.  

\begin{figure}[ht]
\centerline{\epsfig{file=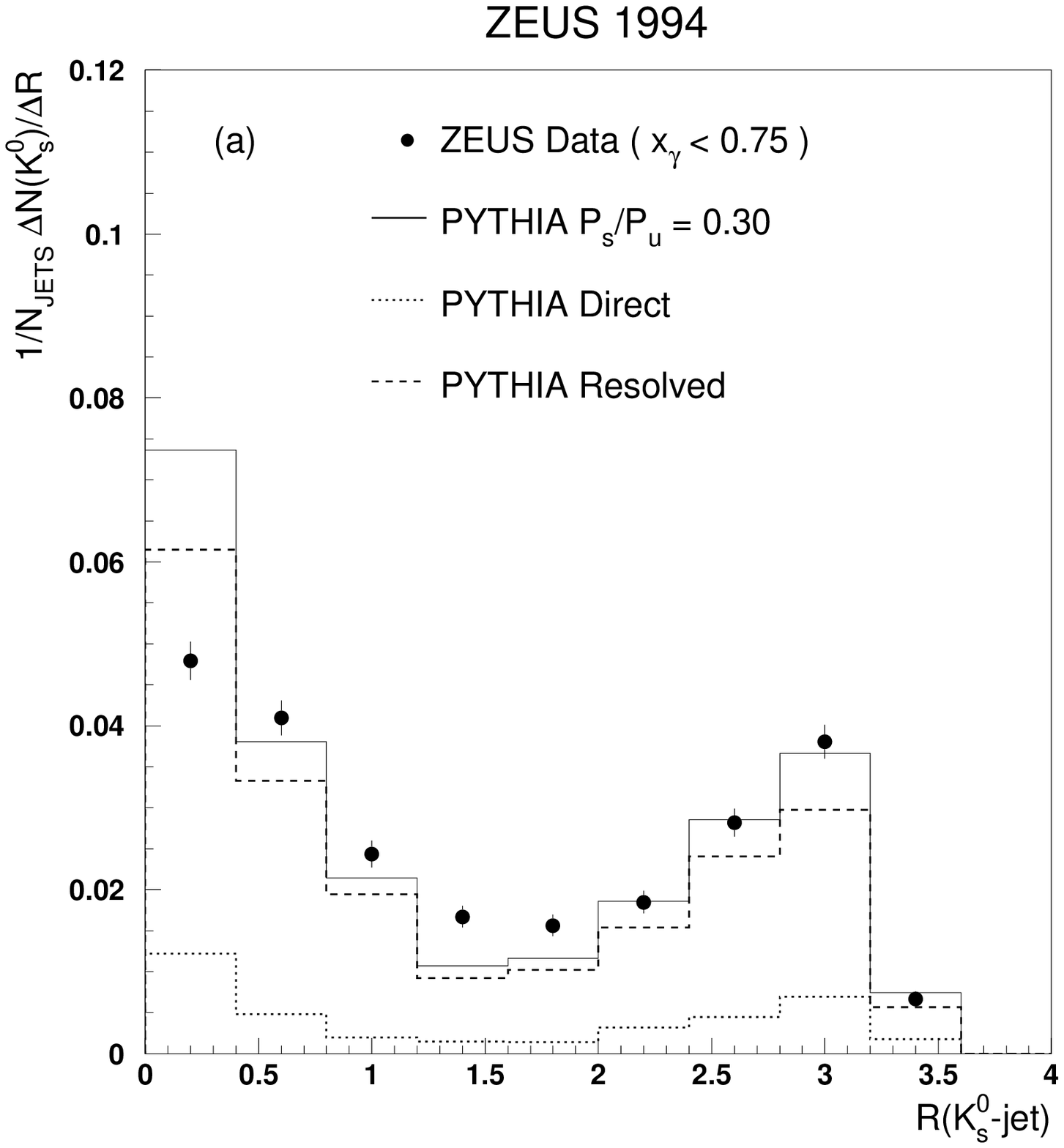,width=10.0cm,%
bbllx=35pt,bblly=196pt,bburx=490pt,bbury=685pt,clip=yes} }
\centerline{\epsfig{file=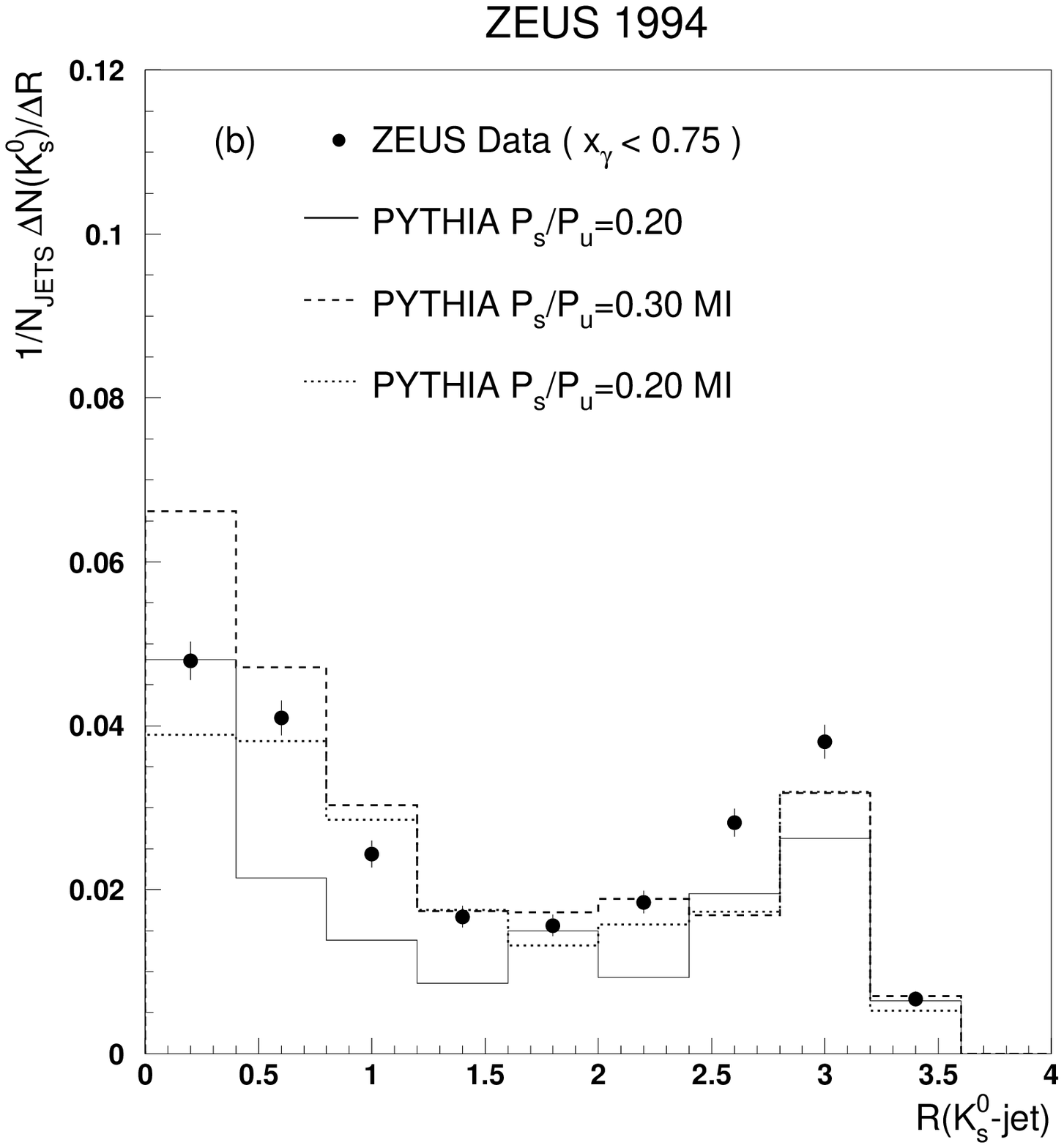,width=10.0cm,%
bbllx=35pt,bblly=196pt,bburx=490pt,bbury=685pt,clip=yes} }
\caption{\footnotesize 
Uncorrected $R$ distributions for resolved-enhanced events (\protect\xgO $<$ 0.75) 
compared (a) with standard version of PYTHIA $(P_s/P_u = 0.3)$, 
  showing also the direct and
resolved contributions separately, (b) with PYTHIA using $ P_s/P_u = 0.2,$  
MI option, and both variants together.  Kinematic definitions are
as in fig.~\protect\ref{ff}.             
\label{frd2}}\end{figure}

\begin{figure}[ht]
\centerline{\epsfig{file=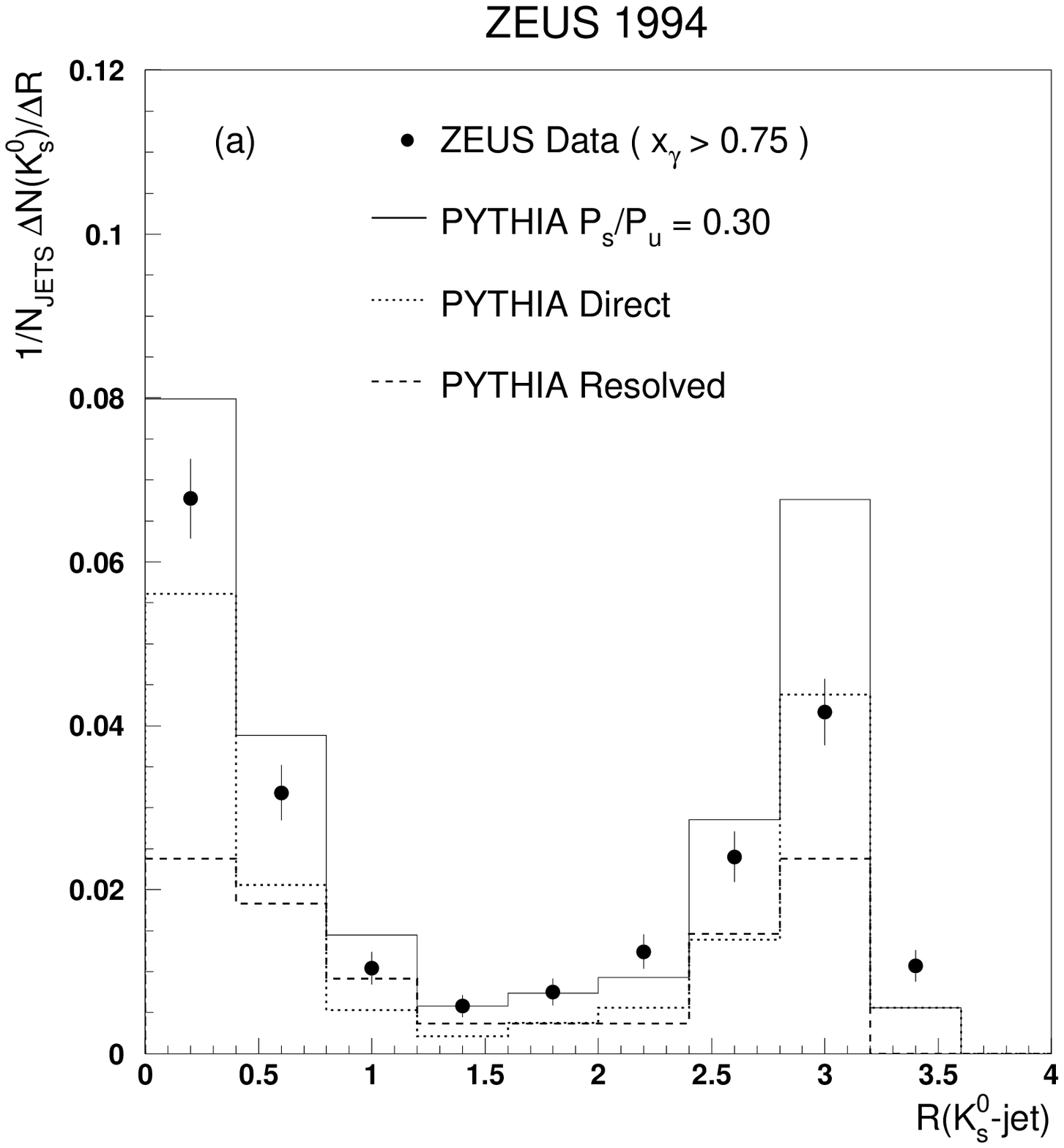,width=10.0cm,%
bbllx=35pt,bblly=196pt,bburx=490pt,bbury=685pt,clip=yes} }
\centerline{\epsfig{file=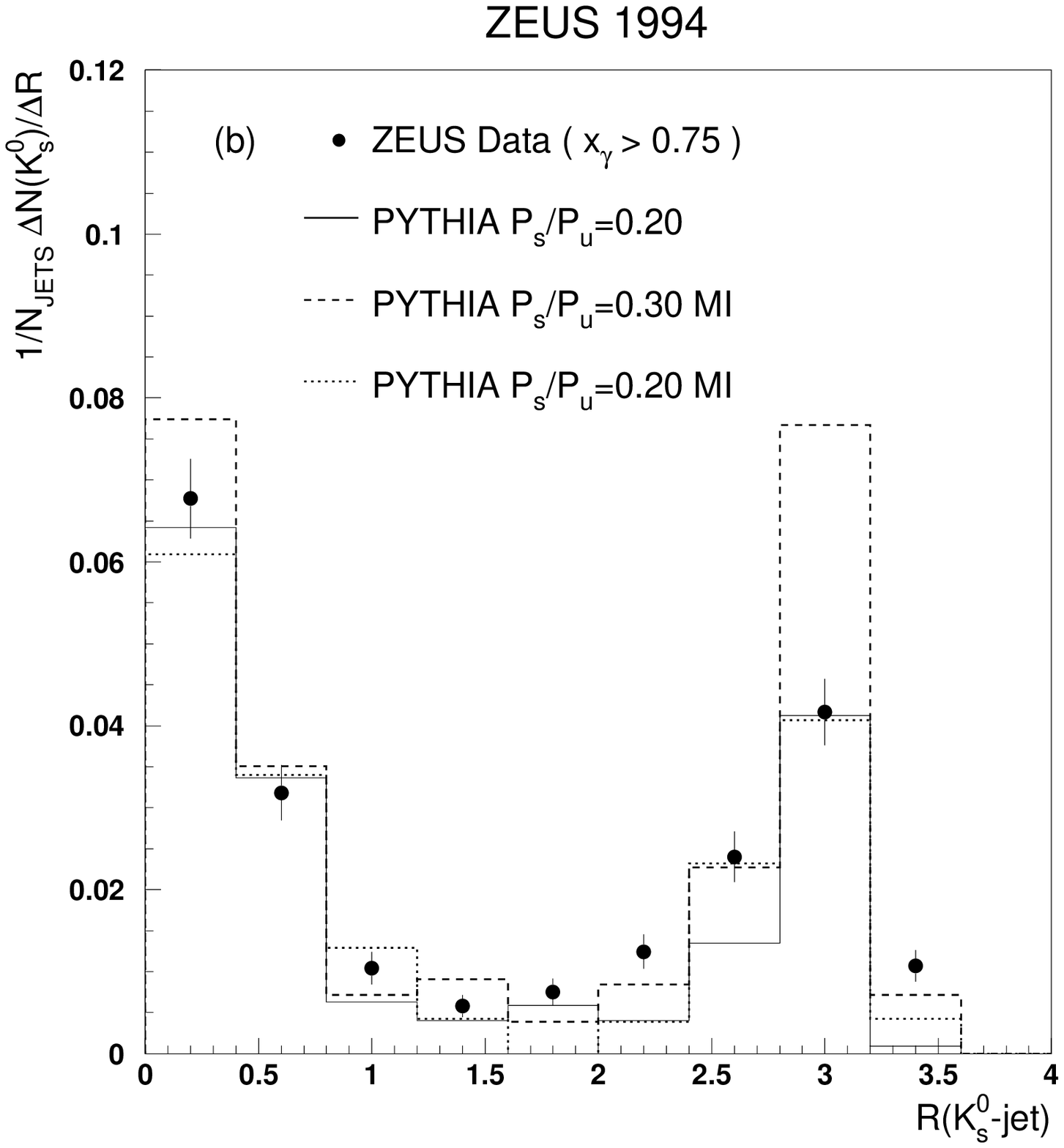,width=10.0cm,%
bbllx=35pt,bblly=196pt,bburx=490pt,bbury=685pt,clip=yes} }
\caption{\footnotesize 
Uncorrected $R$ distributions for direct-enhanced events (\protect\xgO $>$ 0.75) 
compared (a) with standard version of PYTHIA  $(P_s/P_u = 0.3)$, 
showing also the direct and
resolved contributions separately, (b) with PYTHIA using $ P_s/P_u = 0.2,$
MI option, and both variants together.  Kinematic definitions are
as in fig.~\protect\ref{ff}.             
\label{frd1}}\end{figure}

In figs.~\ref{frd2}(a), \ref{frd1}(a), $R$ distributions for the 
resolved-enhanced 
and direct-enhanced data samples ($\xgO < 0.75,\; \xgO > 0.75$) are shown, 
and compared with PYTHIA predictions.  A small admixture of generated 
direct events is found in the PYTHIA samples with $\xgO < 0.75$, and a
relatively larger admixture of generated resolved events 
in the samples with $\xgO > 0.75$.
In the inter-jet region, a deficit in the standard PYTHIA 
predictions is seen in the resolved-enhanced sample 
but not in the direct-enhanced sample, suggesting 
the possibility of an association with multiparton interactions.
In the jet core, the resolved-enhanced sample shows a much larger 
discrepancy between the data and standard PYTHIA than does the direct-enhanced
sample.  The statistical errors on the PYTHIA histograms are similar 
to those on the data. 

In fig.~\ref{frd2}(b), we consider the effects of 
the reduced $P_s/P_u$ value and the MI option in PYTHIA
on the resolved-enhanced event sample.
The use of the MI option is manifestly advantageous,
and is in fact essential in order to obtain reasonable 
agreement with the data when using $P_s/P_u=0.2$. 
A $P_s/P_u$ value between 0.2 and 0.3 appears preferred.
While not perfect, the agreement using  $P_s/P_u = 0.2$ together with the MI 
option represents a significant improvement on the standard version of PYTHIA.
Similar conclusions emerge from the direct-enhanced sample (fig.~\ref{frd1}) 
in which, overall, the reduced $P_s/P_u$ value significantly improves the 
match with the data in the $R$ distribution.  
The MI option, affecting only the resolved contribution to the histograms,
has a positive effect.   A $P_s/P_u$ value 
intermediate between 0.2 and 0.3 might represent a further improvement.

The mean corrected multiplicity of \kz\ with $\pTk \ge 0.5$ GeV 
per jet, integrated over $R\le 1$, is shown in Table~\ref{tbl}.
The full event sample containing one or more jets per event is used. 
A comparison with the different results from PYTHIA
confirms the preference for a lower strangeness suppression parameter
in describing the overall numbers of \kz\ within the jets.

\subsection{Fragmentation functions}
In the previous sections, angular distributions of 
charged particles and \kzs\ relative
to jets were studied, averaged over the particle momenta.   
The momentum distributions of particles within a
jet may be described by fragmentation functions $D(z)$, where $z$ 
is a measure of the fraction of the jet momentum taken by the particle.
$D(z)$ is defined as $(1/N_{jets})\,dN(X)/dz$, where $X$  denotes a  
charged particle or \kz.   A  particle is defined here as being 
associated with a given jet if it satisfies the criterion $R<1$.  

The variable $z$ may be defined in several ways. 
In the usage of CDF~\cite{CDF}, the longitudinal component of
the particle momentum along the jet axis is scaled by the jet energy 
according to a formula which may be written as
$z_L = \pB(X)\cdot\nB(\mbox{jet})/E(\mbox{jet})$. Here 
$\pB$ denotes the momentum 3-vector of a particle, 
and $\nB(\mbox{jet})$ is a unit vector along the jet axis.   
An alternative approach is to take the energy ratio $z_E = E(X)/E(\mbox{jet})$. 
This is analogous to the definition $z = E(X)/E(\mbox{beam})$ normally used 
in $e^+e^-$ collider experiments, where $E(\mbox{beam})$ corresponds to 
the maximum possible energy $E(\mbox{max})$ of a leading parton.  
One notes that in the latter measurements, since they 
are carried out in the $e^+e^-$ centre of mass frame, 
the calculated value of $z$ does not depend on the 
actual production angle of the final state particle, and so is independent 
of many details of the hadronisation mechanism and higher order effects.
No explicit identification of jets is required, or association of particles with jets.  

Here we present our results using both definitions of $z$. 
The numbers of $h^\pm,$ \kz\ and jets were corrected as described in 
section 6.  A small systematic overestimate of the $z$ scale 
at low values exists due to losses of low momentum particles from the
reconstructed jet. The loss of energy in dead material in front 
of the CAL also has the effect of increasing the measured $z$ value.
A correction is applied to remove these effects.  Results are plotted  
at the mean $z$ value of the events in a series of chosen $z$ intervals.
Values of $z_E$ are typically 10--20\% larger on average than those of $z_L$.

\begin{figure}[ht]
\vspace*{-03mm}
\centerline{\epsfig{file=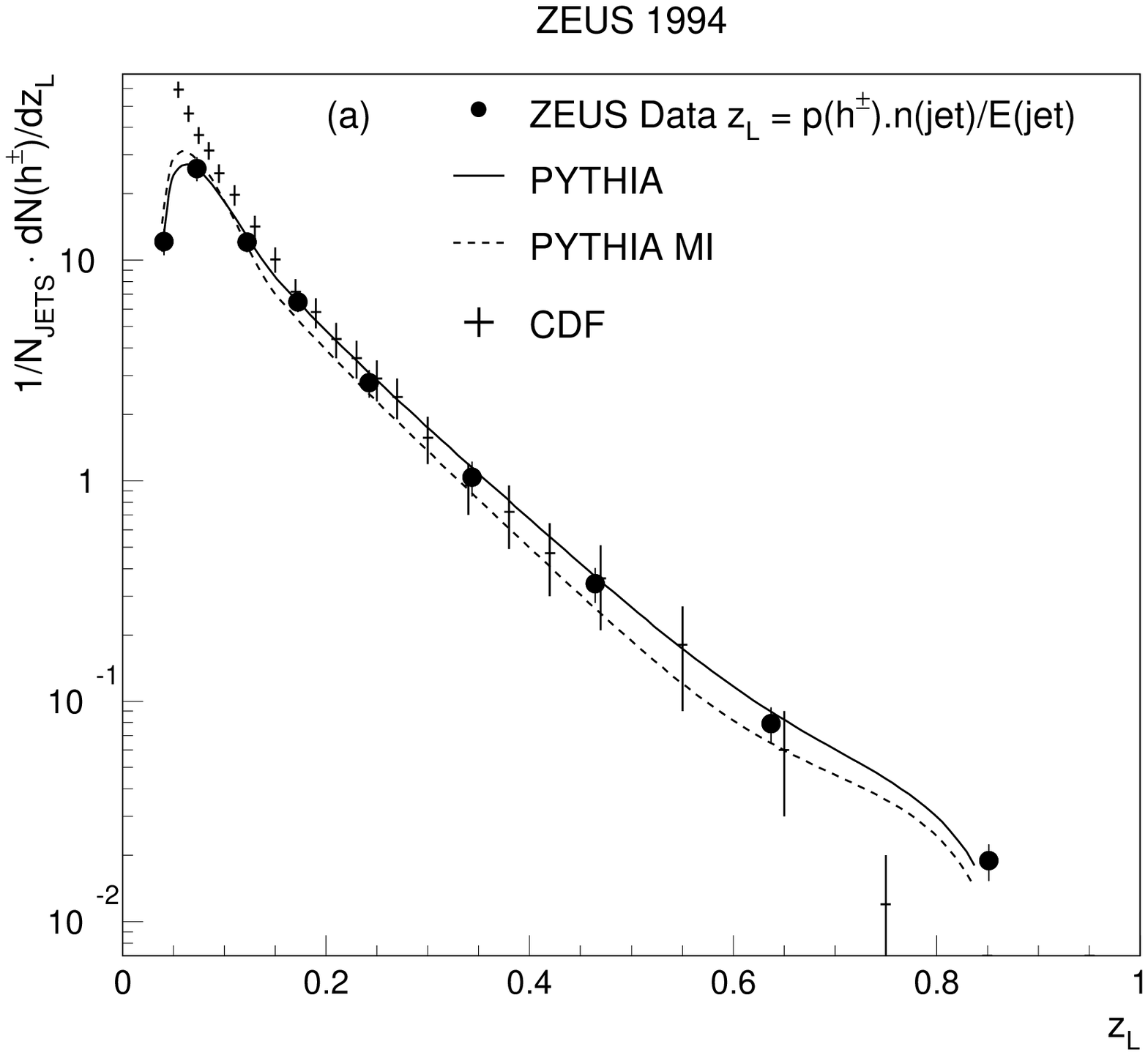,width=11.0cm,%
bbllx=35pt,bblly=190pt,bburx=500pt,bbury=635pt,clip=yes} }
\centerline{\epsfig{file=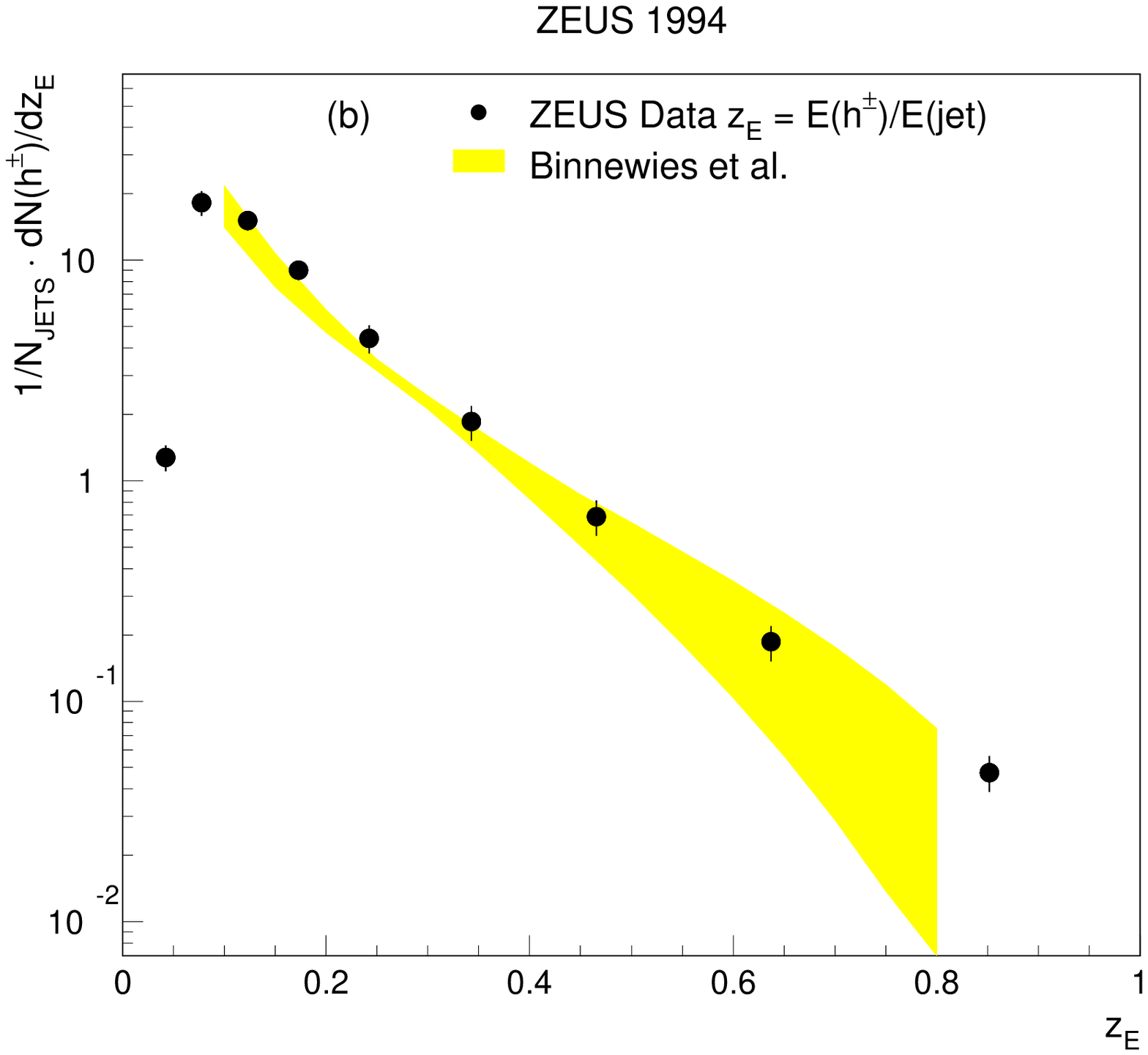,width=11.0cm,%
bbllx=35pt,bblly=190pt,bburx=500pt,bbury=635pt,clip=yes} }
\caption{\footnotesize 
Corrected fragmentation functions $D(z)$ for charged particles $(h^\pm)$ 
with $\pTh\ge 0.5$ GeV, in hadron jets with $\eTj > 8$ GeV and 
$|\etaj|\le 0.5$; systematic errors dominate.
In (a) comparisons are made with variations on the standard version of PYTHIA, and 
with CDF results in the approximate jet energy range 40--100 GeV; 
the dominant systematic errors are shown.  In (b) the data are compared with
a shaded region denoting upper and lower bounds from the phenomenological fit
of \protect\cite{bin}.    
The upper bound corresponds to the fitted  $D(z)$ values for $s$, $d$, $u$ 
quarks, respectively, in the intervals $z$ = 0.1--0.25, 0.25--0.45,
0.45--0.8.   The lower bound corresponds similarly to $u$, $c$ quarks in the 
intervals $z$ = 0.1--0.35, 0.35--0.8, respectively. 
\label{fD1}}\end{figure}

\begin{figure}[ht]
\centerline{\epsfig{file=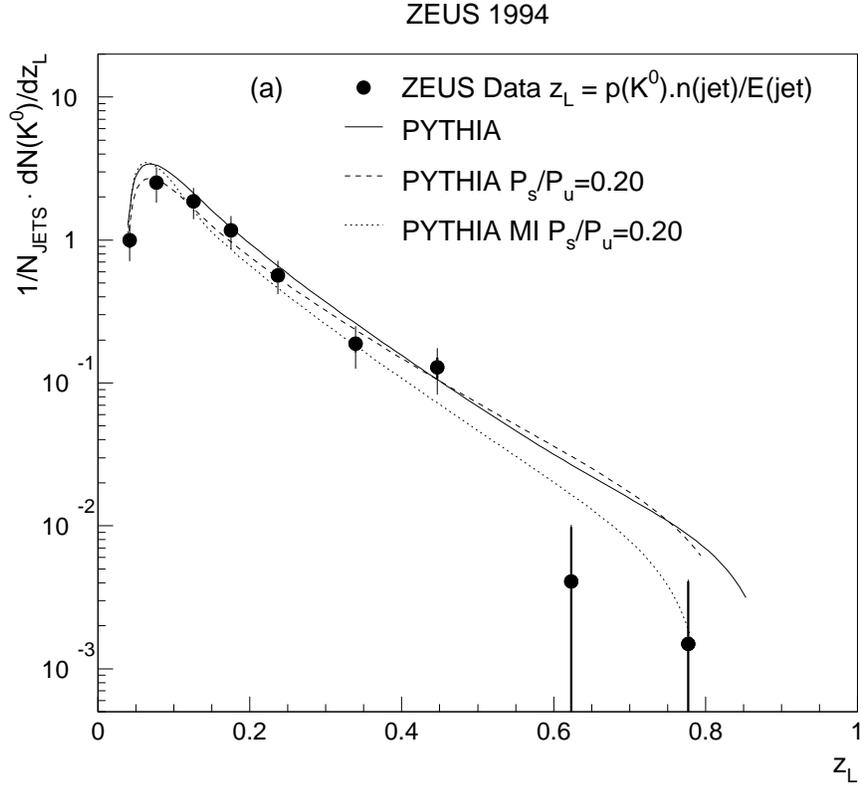,width=11.0cm,%
bbllx=35pt,bblly=190pt,bburx=500pt,bbury=635pt,clip=yes} }
\centerline{\epsfig{file=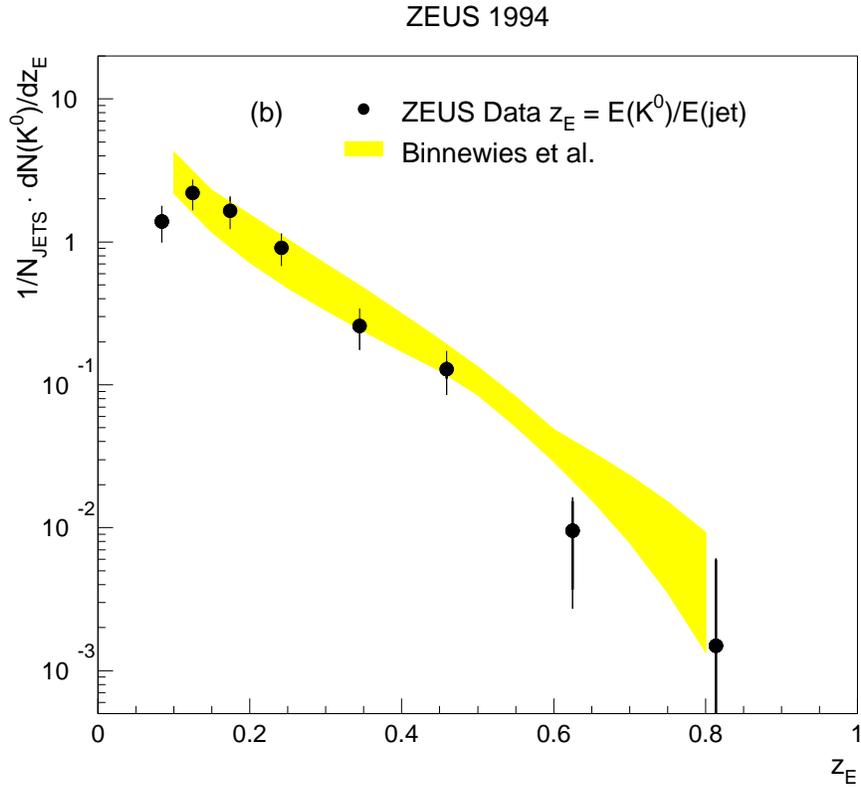,width=11.0cm,%
bbllx=35pt,bblly=190pt,bburx=500pt,bbury=635pt,clip=yes} }
\caption{\footnotesize 
Corrected fragmentation functions $D(z)$ for \kz.  
Details as for previous figure.  The upper bound of the
shaded region in (b) corresponds to fitted $D(z)$ values from \protect\cite{bin}
for $d(=s)$, $c$, $u$ quarks in the intervals $z$ = 0.1--0.2, 0.2--0.45, 0.45--0.8,
respectively. The lower bound corresponds to  $g$, $d(=s)$ in 
the intervals $z$ = 0.1--0.45, 0.45--0.8, respectively. 
Thick error bars denote dominant statistical errors, thin denote dominant
systematic errors. 
\label{fD2}}\end{figure}

Figs.\ \ref{fD1}(a) and \ref{fD2}(a) show the resulting $D(z_L)$ 
distributions for $h^\pm$  and \kz\ respectively,
with pion masses assigned to the \hpm.  Also plotted are values 
of the corresponding distribution obtained using 
PYTHIA.  The distributions are averaged over $\etaj$ and
$\eTj$ within the same selected ranges as above.  
For $z_L<0.05$ the effects of the 0.5 GeV
cut-off on $p_T$ in the \hpm\ and \kz\ data become important. 
Good agreement is seen between the 
data and the standard PYTHIA distributions,
in the \kz\ case for $z_L$ values above 0.15; 
in this range it is not possible to discriminate between  
$P_s/P_u$ values of 0.3 and 0.2.  
The discrepancy noted in the  \kzs\  $R$ plots appears to be concentrated 
at $z_L < 0.15$, where the bulk of the statistics lie, apart from the
two highest $z_L$ points with large errors.  
The MI is in agreement with the $\hpm$ data at low momenta now that the
association of particles with jets has been made.  
Also shown in fig.\ \ref{fD1}(a) are $D(z_L)$ values from CDF~\cite{CDF}, 
for \hpm\ in jets of energy $\approx$40--100 GeV produced in $p\bar p$ 
scattering.  Although the CDF energies are considerably higher than those of the 
present measurements, and the mixture of leading final state partons is also 
likely to be different, there is a clear similarity of 
the fragmentation functions in the range $0.15 < z_L < 0.7$.

For a  comparison with $e^+e^-$ data, results from the
present measurements are first compared with calculations from 
the next-to-leading order phenomenological fits of Binnewies et al~\cite{bin}
(figs.\ \ref{fD1}(b),  \ref{fD2}(b)).  These authors
have made use of a variety of $e^+e^-$ data sets at different energies 
to extract separate fragmentation functions for the different 
types of primary parton in the range $0.1 < z < 0.8$.
The approach is based on the different quark couplings to 
photons and  to $Z$ bosons, and on topological properties of 
events with hard gluon emission.  A definition 
$z = E(X)/E\mbox{(beam)}$ is used; we therefore adopt here the corresponding 
definition $z = z_E = E(X)/E\mbox{(jet)}$.
Since the primary partons in the photoproduced jets are unidentified, 
and we currently lack a full NLO Monte Carlo simulation of
the $\gamma p$ process, a shaded region is 
drawn  to cover the spread of the calculated $D(z)$ values for the fragmentation 
of gluons and different types of quark (i.e.\ $u$, $d$, $s$, $c$).
The calculation has been performed using a QCD scale of 8 GeV,
corresponding approximately to the present jet \eT\ values.
Details of which parton fragmentation functions contribute to the
upper and lower bounds of the shaded regions are given in the figure captions.  

Good agreement with the calculations is found, to within their 
uncertainty as applied to the present data.  A few remarks need to be 
held in mind:
\begin{romlis}
\item Ref.~\cite{bin} includes charged decay products of long-lived primaries 
such as \kzs, $\Lambda$ in the definition of a charged particle.  This gives 
$\approx10$\% more charged hadrons compared with the present method.
\item As stated above, a jet definition, and the requirement that the given particle 
be associated with a jet, are required in our present approach, 
but not in the  $e^+e^-$ based analyses.
This tends to lower the distributions in particular at lower $z_E$ values.
\item The experimental  $z_E$ (or $z_L$) value is corrected relative to the energy of
hadron jets. A correction to the energy of the 
the final-state leading parton in the hard process has not been attempted here.
\end{romlis}

More precise comparisons are available using the direct photoproduction 
process, since a close similarity is expected in general between event 
properties in  direct hard photoproduction, $e^+e^-$ annihilation, 
and DIS at low $x$, where $x$ is the Bjorken variable.  At energies 
comparable to those of the present data, 
$e^+e^-$ annihilation is governed by a single electromagnetic vertex, 
summed over the different types of quark, while the LO direct 
photoproduction process is
dominated by photon gluon fusion, whose diagrams feature 
a similar electromagnetic vertex accompanied by a quark-gluon vertex which 
does not depend on the quark type.  In a similar way, in DIS events at low $x$, 
the virtual photon couples to quarks that come predominantly from the sea,
having evolved through $gq\bar q$ vertices which are likewise flavour
independent. 

\begin{figure}[ht]
\centerline{\epsfig{file=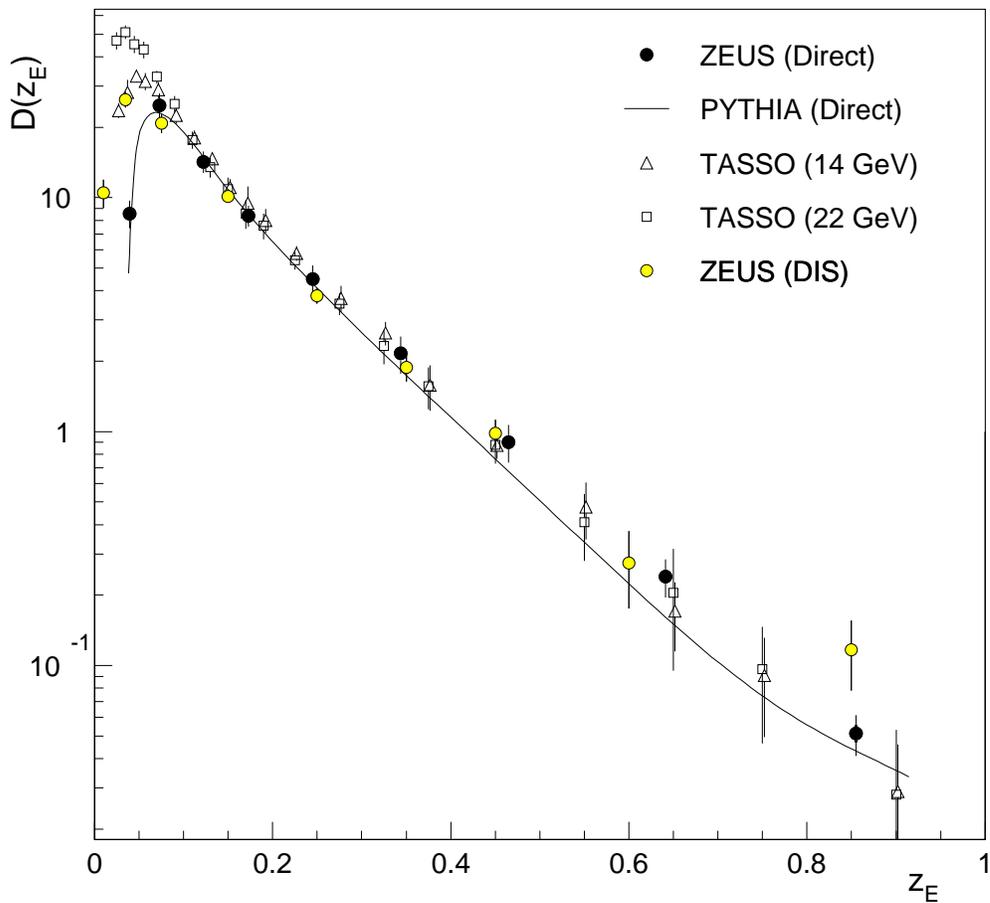,width=15.0cm}}
\caption{\footnotesize 
Fragmentation functions $D(z_E)$ for $h^{\pm}$, obtained from direct-enhanced
events and corrected to ``pure direct" values.  Also plotted are
(i) standard PYTHIA predictions for the LO direct process,
(ii) data from $e^+e^-$ annihilation at
similar jet energies with centre of mass energies as
indicated~\protect\cite{TASSOh}, halved to take account of the dominant 
$q\bar q$ final state, and (iii) data from ZEUS measured in deep inelastic 
scattering~\protect\cite{ZEUSDIS} with Breit frame jet energies in the 
range 6.3-8.9 GeV.   
The variable $z_E$ is defined as $E(\hpm)/E\mbox{(jet)}$ for ZEUS 1994 
photoproduction data and  $z_E = E(\hpm)/E\mbox{(max)}$ for the other
experimental points. 
Statistical and total errors on the ZEUS 1994 data are shown 
as for previous figures; errors 
are total errors for the TASSO and ZEUS DIS points.  
\label{fzzh}}\end{figure}

\begin{figure}[ht]
\centerline{\epsfig{file=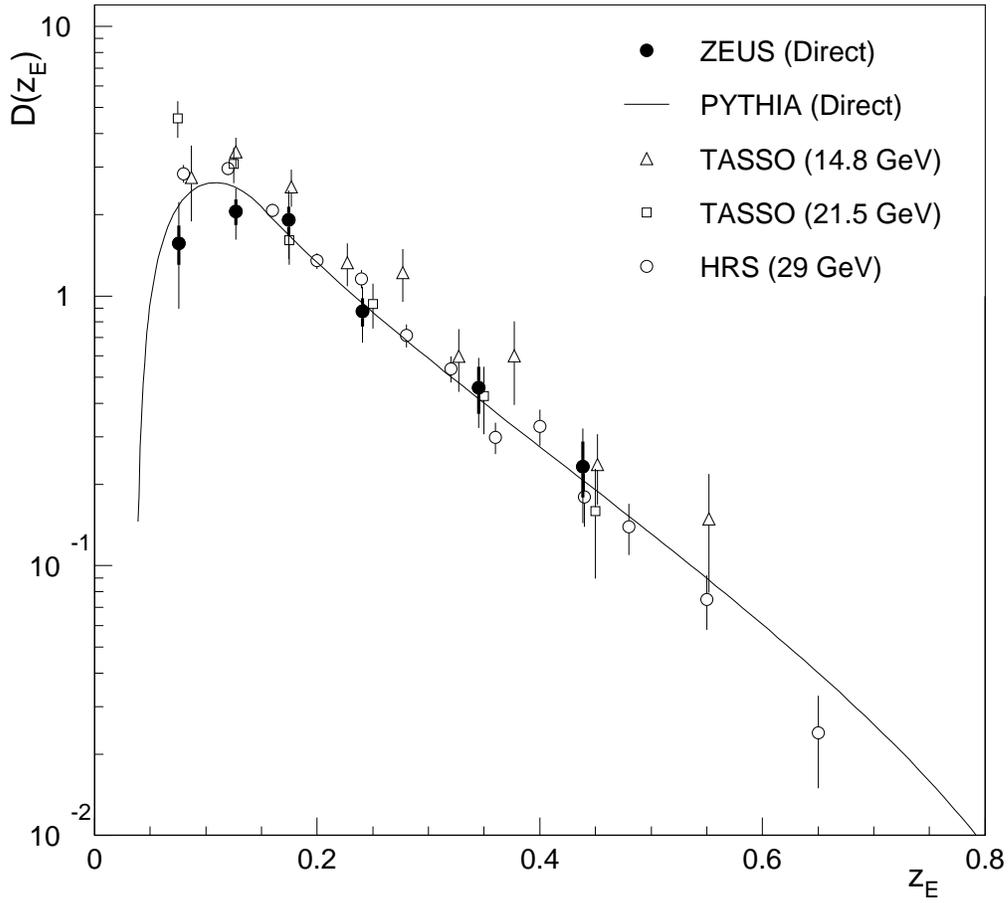,width=15.0cm}}
\caption{\footnotesize 
Fragmentation functions $D(z_E)$ for \kz, obtained from direct-enhanced
events and corrected to ``pure direct" values;  
$z_E$ definitions correspond to those for \hpm. Also plotted are
standard PYTHIA predictions for the LO direct process and data from 
previous experiments at similar jet energies, with centre of mass energies as 
indicated~\protect\cite{TASSO,HRS}.  Errors on the TASSO and HRS points are total
errors.  
\label{fzz}}\end{figure}

We therefore make a comparison between fragmentation functions obtained from 
the direct photoproduction process in ZEUS, and values obtained from $ e^+e^-$
and DIS measurements.  
Figs.~\ref{fzzh}, \ref{fzz} show results obtained from the present 
direct-enhanced event samples, corrected by means of PYTHIA on a bin-by-bin 
basis to remove the resolved component so as to be  
equivalent to results from ``pure direct" event samples.
These are compared with predictions from PYTHIA, and with $e^+e^-$ 
measurements  at similar centre of mass energies to those 
of the present hard subprocess, i.e.\ twice the transverse energy of the 
selected jets.  A factor of 0.5 is applied to the published $e^+e^-$
data to allow for the dominant $q\bar q$ final state.

For the $h^\pm$ data, a further comparison is made to
fragmentation functions from ZEUS calculated in the current region of the
Breit frame in deep inelastic scattering~\cite{ZEUSDIS}.  Here, 
data points are used covering a photon virtuality range of 
$160 < Q^2 < 320$ GeV$^2$ with  $0.0024 < x < 0.01$. 
The corresponding jet energy is $Q/2$, i.e.\ 6.3--8.9 GeV, roughly 
equivalent to the jets of the present events.
It should be noted that the TASSO data~\cite{TASSOh} include
all charged products from \kzs\ and $\Lambda$ decays, while the ZEUS DIS
and ZEUS 1994 photoproduction data exclude all such products; 
these effects contribute at the level of $\pm10\%$.   
The \hpm\ data compare well also with $pp$ data from the ISR (not 
shown)~\cite{ISR} although a different parton mixture is expected in the final
state here. 
For the inclusive \kz\ data of \cite{TASSO,HRS} at different fixed centre of 
mass energies, the authors quote scaling cross sections which 
have been converted to fragmentation functions to compare with the present data. 

At low $z_E$, the distributions may be affected by differences 
between the colour flows in the different types of event, as well as 
by the cut-off imposed at 0.5 GeV in \pT\ in the present
$h^\pm$ and \kz\ data, making comparisons between the different data sets 
difficult.  However for $z_E > 0.1$ for \hpm, and $z_E > 0.15$ for \kz, 
the present results are in good agreement with the standard PYTHIA
predictions, and the $D(z_E)$ values from the different 
measurements  are also in good agreement with each other.
This  well illustrates the universality that is believed to be
a property of the quark fragmentation process.  

\section{Summary and conclusions} 

We have studied the properties of charged particles (\hpm) and \kz\ mesons in 
photoproduced events in the ZEUS detector at HERA.  
The \kz\ mesons were studied in the $\kzs\to \pi^+\pi^-$ decay mode.
In each event at least one reconstructed jet was required 
in the calorimeter with measured $\eTrec > 7$ GeV, 
centrally produced in the laboratory frame.  
The distributions of the $h^\pm$ and \kz\ in these events were 
studied as a function of a number of kinematic variables; 
the distance $R$ in $(\eta,\phi)$ of the particle from the axis of a jet  
displayed information which 
was not so clearly evident from the other distributions.

Correction factors were applied to evaluate the numbers of \hpm\ and \kz\ 
per jet at the final state hadron level, with $\eTj > 8$ GeV 
and $|\etaj| < 0.5$, for events in the $\gamma p$ centre of mass energy
range $134 < W < 277$ GeV.  Corrected distributions in 
transverse momentum \pT\ and pseudorapidity $\eta$ are given.  
The corrected numbers of \hpm\ and \kz\ within a jet are evaluated,
with a particle defined as being inside a jet if it is within unit 
radius of the jet axis in $(\eta, \phi)$. 
Fragmentation functions $D(z)$  for \hpm\ 
and \kz\ in photoproduced jets have also been determined.
In comparing the present results with those from theory and from other
experiments, two definitions of $z$ are employed, in which the longitudinal 
momentum component of the particle along the jet axis is scaled to the 
jet energy ($z_L$), or the particle energy is similarly scaled ($z_E$).

The distribution of \hpm\ within photoproduced jets
is found to be fairly well described by the standard version of PYTHIA, 
whereas that of \kz\ mesons is not. Outside the jets, more \hpm\   
and \kz\ are found than predicted; the latter situation  can be to some
extent remedied by using a version of PYTHIA which 
includes a simulation of multiparton interactions in resolved events. Taken 
overall, the numbers of \kz\ within the jets correspond to a reduced value 
of the strangeness suppression parameter $P_s/P_u$ in PYTHIA. 
When the data are divided into samples enriched in the resolved and
direct photoproduction processes respectively, this statement remains true 
in either case.  However the effect is concentrated at $z$ values below
$\approx0.15$; 
at higher values the default parameters of
PYTHIA give a satisfactory description of the data.  This suggests a need for
further study of fragmentation at low $z$, where large numbers
of particles are found but where their association with jets may be less well
defined than at high $z$.

The fragmentation functions determined using  
inclusive photoproduced jets are found to be in agreement 
with calculations from Binnewies et al.\ to within the uncertainty 
of the calculation as applied to the present data. 
A close similarity is seen between the present \hpm\ fragmentation functions 
in the range $0.15 < z_L < 0.7$ and those observed in $p\bar p$
scattering.  Fragmentation functions have  also been extracted for 
\hpm\ and \kz\ in direct photoproduced jets, and  are compared with 
corresponding data from $e^+e^-$  annihilation  and 
from deep inelastic scattering.  Agreement is 
good for $z_E > 0.1$ and $z_E > 0.15$ for \hpm\ and \kz\ respectively.
This, together with the agreement found in this region with 
PYTHIA, represents a confirmation of the idea of a universally 
valid description of parton fragmentation.
\\[8mm]

\noindent
{\Large \bf Acknowledgements}\\[3mm]
We thank the DESY directorate and staff for their continued 
support and encouragement, and likewise the HERA
machine group for their excellent efforts in operating HERA. 
We are grateful to J. Binnewies for helpful conversations and for making
numerical calculations available.

\vfill
\begin{table}[h] 
\centerline{\begin{tabular}{|l|c|llll|} 
\hline  \rule[1.70ex]{0ex}{1ex}
 & DATA & \multicolumn{4}{c|}{PYTHIA Monte Carlo\rule[-0.5ex]{0ex}{1ex}} \\
 & (ZEUS 1994) &  Standard & MI & Standard & MI \\ \hline
\rule[1.5ex]{0ex}{1ex}$P_s/P_u$ & & 0.3 & 0.3 & 0.2 & 0.2 \\
\hline
\rule[1.8ex]{0ex}{1ex}$h^\pm$   & 3.25$\pm$0.02$\pm$0.28 & 
 3.31$\pm$0.03 & 3.44$\pm$0.06 & 3.37$\pm$0.04 & 3.46$\pm$0.04 \\
 $K^0$ & 0.431$\pm$0.013$\pm$0.088 &
 0.513$\pm$0.013 & 0.560$\pm$0.025 & 0.416$\pm$0.014 & 0.426$\pm$0.013 \\
\hline\end{tabular}}
\vspace*{3mm}
\caption{
\footnotesize Corrected multiplicities of charged particles ($h^\pm$) and neutral
$K^0$ mesons ($\pT \ge 0.5$ GeV) per photoproduced jet ($\eTj > 8$ GeV,
$|\etaj|<0.5$), cone radius = 1, for $134 < W < 277$ GeV.
Comparison is made between ZEUS data and PYTHIA 
Monte Carlo predictions with strangeness
suppression parameter ($P_s/P_u$) = 0.3 (standard value) and 0.2, without and with
the multiparton interactions (MI) option.  The first error quoted is in
each case statistical, the second error on the data is systematic.  
\label{tbl}}\end{table}
\vfill
\newpage

\end{document}